\documentclass[useAMS,usegraphicx]{mn2e}\usepackage[totalwidth=450pt,totalheight=650pt,a4paper]{geometry}

\newcommand{\et}{\sl et al. \rm}

\title[The UKIRT Infrared Deep Sky Survey (UKIDSS)]{The UKIRT Infrared Deep Sky
 Survey (UKIDSS)
 }

\author[Lawrence et al.]{
 A.Lawrence$^{1}$, 
S.J.Warren$^{2}$, 
O.Almaini$^{3}$, 
A.C.Edge$^{4}$,
N.C.Hambly$^{1}$, \newauthor
R.F.Jameson$^{5}$,
P.Lucas$^{6}$, 
M.Casali$^{7}$,
A.Adamson$^{8}$, 
S.Dye$^{9}$, 
J.P.Emerson$^{10}$, \newauthor
S.Foucaud$^{3}$, 
P.Hewett$^{11}$, 
P.Hirst$^{8}$, 
S.T.Hodgkin$^{11}$, 
M.J.Irwin$^{11}$, \newauthor 
N.Lodieu$^{5}$, 
R.G.McMahon$^{11}$  
C.Simpson$^{12,13}$, 
I.Smail$^{4}$,
D.Mortlock$^{2}$,
M.Folger$^{7}$  \newauthor \\
$^{1}$Institute for Astronomy, SUPA (Scottish Universities Physics Alliance), University of Edinburgh, \\ 
Royal Observatory, Blackford Hill, Edinburgh EH9 3HJ \\
$^{2}$Blackett Laboratory, Imperial College of Science Technology
and Medicine, Prince Consort Rd, London SW7 2AZ \\
$^{3}$School of Physics and Astronomy, University of Nottingham,
University Park, Nottingham NG7 2RD\\
$^{4}$Institute for Computational Cosmology, Durham University, South
Road, Durham DH1 3LE\\ 
$^{5}$Department of Physics and Astronomy, University of Leicester,
Leicester LE1 7RH\\ 
$^{6}$Centre for Astrophysics Research, Science and Technology
Research Institute, University of Hertfordshire, Hatfield, AL10 9AB \\
$^{7}$UK Astronomy Technology Centre,Royal Observatory,
Blackford Hill, Edinburgh EH9 3HJ  \\
$^{8}$Joint Astronomy Centre, 660 N. A'ohuku Place, Hilo, Hawaii
96720, U.S.A. \\ 
$^{9}$School of Physics and Astronomy, Cardiff University, 5 The Parade, 
Cardiff, CF24 3YB \\
$^{10}$Astronomy Unit, School of Mathematical Sciences, Queen Mary
University of London, Mile End Road, London E1 4NS \\
$^{11}$Institute of Astronomy, University of Cambridge, Madingley
Road, Cambridge CB3 OHA\\ 
$^{12}$Department of Physics, Durham University, South Road,
Durham DH1 3LE\\
$^{13}$Astrophysics Research Institute, Liverpool John Moores University, Twelve
Quays House, Egerton Wharf, Birkenhead CH41 1LD\\
}

\begin{document}

\date{Accepted 2007 May 29. Received 2007 May 28 ; in original form 2006 April 7 }

\pagerange{1599--1617} \volume{379} \pubyear{2007}

\maketitle

\label{firstpage}

\begin{abstract}

We describe the goals, design, implementation, and initial progress of the UKIRT Infrared Deep Sky Survey (UKIDSS), a seven year sky survey which began in May 2005. UKIDSS is being carried out using the UKIRT Wide Field Camera (WFCAM), which has the largest {\em \'{e}tendue} of any IR astronomical instrument to date. It is a portfolio of five survey components covering various combinations of the filter set ZYJHK and H$_2$. The Large Area Survey, the Galactic Clusters Survey, and the Galactic Plane Survey cover approximately 7000 square degrees to a depth of K$\sim$18; the Deep Extragalactic Survey covers 35 square degrees to K$\sim$21, and the Ultra Deep Survey covers 0.77 square degrees to K$\sim$23. 
Summed together UKIDSS is 12 times larger
in effective volume than the 2MASS survey. 
The prime aim of UKIDSS is to provide a long term astronomical legacy database; the design is however driven by a series of specific goals -- for example to find the nearest and faintest sub-stellar objects; to discover Population II brown dwarfs, if they exist; to determine the substellar mass function; to break the z=7 quasar barrier; to determine the epoch of re-ionisation; to measure the growth of structure from z=3 to the present day; to determine the epoch of spheroid formation; and to map the Milky Way through the dust, to several kpc. The survey data are being uniformly processed. Images and catalogues are being made available through a fully queryable user interface - the WFCAM Science Archive (WSA : http://surveys.roe.ac.uk/wsa). The data are being released in stages. The data are immediately public to astronomers in all ESO member states, and available to the world after eighteen months. Before the formal survey began, UKIRT and the UKIDSS consortium collaborated in obtaining and analysing a series of small science verification (SV) projects to complete the commissioning of the camera. We show some results from these SV projects in order to demonstrate the likely power of the eventual complete survey. Finally, using the data from the First Data Release we assess how well UKIDSS is meeting its design targets so far.
\end{abstract}



\section{Introduction}  \label{intro}

The UKIRT Infrared Deep Sky Survey (UKIDSS) 
can be considered the near-infrared counterpart of the 
Sloan Digital Sky Survey (SDSS; York \et 2000).
It does not
cover the whole sky, but is many times deeper than the Two Micron All Sky Survey 
(2MASS; Skrutskie \et 2006).
 It is in fact
not a single survey but a survey programme combining a set of five
survey components of complementary combinations of depth and
area, covering several thousand square degrees to $K\sim 18$, 35 square
degrees to $K\sim 21$, and 0.77 square degrees to $K\sim 23$. The survey uses
the Wide Field Camera (WFCAM) on the 3.8m United Kingdom Infra-red Telescope
(UKIRT). WFCAM has an instantaneous field of view of 0.21 square degrees,
considerably larger than any previous IR camera on a 4m class telescope, 
along with a pixel size of 0.4\arcsec. 
The tip-tilt system on UKIRT delivers close to natural seeing (median size 0.6\arcsec) across the whole field of view.
The combination of large telescope, large field of view, and good
image quality, makes possible a survey of considerably greater scope
than 2MASS.
The various
surveys employ up to five filters $ZYJHK$ covering the wavelength range
$0.83-2.37\,\mu$m and extend over both
high and low Galactic latitudes. 
The survey began on 2005 May 13, and is expected to take seven
years to complete. The survey is being carried out by a private consortium but
is fully public with no proprietary rights for the consortium.
Data products are being being released in stages, with the intention
of having these roughly twice a year.

\subsection{Origins and nature of project}  \label{origins}

The UKIDSS survey concept first emerged in 1998 while making the funding case
for the WFCAM instrument itself, but eventually became a formal refereed
proposal to the UKIRT Board in March 2001, submitted by a consortium of 61 UK
astronomers. This included a commitment to making data available immediately to
all UK astronomers (not just consortium members), 
plus specified individual Japanese consortium members, and
available to the world after a year or two. (See  section \ref{releases} for
the final data release policy.) Later, during the UK's entry in to ESO, it was
agreed that astronomers in all ESO member states would have the same data rights
as UK astronomers, and at the same time, membership of the consortium was
extended to any interested European astronomers. Consortium membership now
stands at 130. Note that individual astronomers are members, not their
institutions.

The project is unusual compared to previous large survey projects, being
neither private, nor conducted by a public body on behalf of the community.
UKIDSS relies on the separate existence of three things. (i) The UKIRT
observatory, operated as part of the UK's Joint Astronomy Centre (JAC). (ii) The
WFCAM instrument, built at the Astronomy Technology Centre (ATC) at the Royal
Observatory Edinburgh, as a funded PPARC project. Note that WFCAM was built as a
common user instrument to be part of the UKIRT suite of instruments. The UKIDSS
consortium is essentially the largest single user. (iii) The pipeline and
archive development project, run by the Cambridge Astronomy Survey Unit (CASU)
and the Edinburgh Wide Field Astronomy Unit (WFAU), and funded by several
different PPARC grants. This data processing development is part of the VISTA
Data Flow System (VDFS) project, with the WFCAM pipeline and archive being seen
as an intermediate step. Note this data processing project deals with all
WFCAM data, not just the UKIDSS data.

The consortium has no data privileges, and does not plan the scientific exploitation
of the survey; its purpose is to make the survey happen, on behalf
of the ESO communuity.
The aim of the UKIDSS consortium is then to produce the scientific design for
the survey; to win the telescope time necessary; to plan the implementation of
the survey, liaising with the other bodies above; to staff the
observing implementation; to define the necessary Quality Control (QC) filtering stages 
to produce final survey products; to assist the data processing team as necessary
in producing stacked and merged survey products; and finally to document the
production of the survey data in scientific publications and other technical papers.
 A number of individuals in ATC, JAC, CASU and WFAU are
also members of UKIDSS, so that the liaison with the camera construction and
data processing projects, as well as telescope operations, has been well
motivated. A clear relationship with UKIDSS has been built into each of these
projects. For example the science requirements document for the pipeline and
archive emerged from consultation with UKIDSS; and the commissioning schedule
for WFCAM include a ``science verification (SV)'' phase following standard tests, in
which the survey implementation described in section \ref{WFCAM-imp} could be tested
and refined.

\subsection{Technical reference papers}  \label{techpapers}

This paper is one of a set of five which
provide the reference technical documentation for UKIDSS.
It summarises the scope, goals, and overall design of the survey,
along with a brief discussion of implementation methods, progress
to date, and 
presentation of early ``science verification'' data.
The other four papers, described briefly
below, are Casali \et  (2007), Hewett \et  (2006), Irwin \et
(in preparation) and Hambly \et  (in preparation). In addition to these five core reference
papers, each data release will be accompanied by a paper detailing 
its contents and implementation information. The first two of these are 
Dye \et (2006) for 
the ``Early Data Release (EDR)'', and Warren \et (2007) for the ``First Data 
Release'' (DR1).

Casali \et (2007) describe the survey instrument, WFCAM. A short summary 
is given in section \ref{WFCAM-imp}. At the time of commissioning, 2004
November, the instrument {\em \'{e}tendue}\footnote{product of telescope collecting
area, and solid angle of instrument field of view, sometimes called {\em grasp}}
of $2.38\,$m$^2\,$deg$^2$ was the largest of any near-infrared imager in
the world. The Canada France Hawaii Telescope WIRCam instrument (Puget \et
2004) covers a solid angle of 0.1$\,$deg$^2$ per exposure giving an {\em
\'{e}tendue} of $1.11\,$m$^2\,$deg$^2$.  WFCAM is likely to remain as the
near--infrared imager with the largest {\em \'{e}tendue} in the world until completion
of the near--infrared camera for VISTA (Dalton \et  2004).

The data flow system for WFCAM is described by Irwin \et (in preparation) and Hambly \et (in preparation). A
summary is given in section \ref{sectn:data}. The very high data rate (1TB per week) requires 
a highly automated processing system that removes instrumental signature, produces object catalogues, and ingests into a fully queryable WFCAM Science Archive (WSA). It is expected
that nearly all science analysis of UKIDSS will be initiated through the WSA.

The photometric system is described in Hewett \et (2006).
The survey uses five broadband filters,  $ZYJHK$. The $JHK$ passbands are as
close as possible to the MKO system; the $Z$ passband is similar to the SDSS
$z'$ passband, but has a cleaner red tail. The $Y$ passband is a new one centred
at $0.97\,\mu$m, which bridges the gap between Z and J. The exact passband was 
designed with the aim of discriminating between high-redshift
quasars and brown dwarfs.

Hewett \et present the measured passband
transmisssions, and use synthetic colours of various classes of astronomical
object to produce expected colour equations between certain WFCAM, SDSS, and
2MASS filters. Table \ref{passbands} provides summary information on 
the ZYJHK passbands.
A later paper (Hodgkin \et in preparation) will report on the photometric  
calibration of the UKIDSS survey and colour equations determined on  
the sky from standard star observations.


\begin{table}
\caption{\it WFCAM passbands used for UKIDSS. 
Columns show the name, the effective wavelength, 
(Schneider, Gunn \& Hoessel, 1983), the short and long
wavelength limits, and the width, i.e. the difference between these two wavelengths.
The wavelength limits are set by the total system transmission, and
defined as 50\% transmission relative to the peak. Complete
transmission functions are shown in Hewett et al (2006). 
}
\label{passbands}
\centering
\begin{scriptsize}
\begin{tabular}{cccc}\\ \hline
band & $\lambda_{\mathrm{eff}}$ & range  & width \\
      &   $\mu$m        &   $\mu$m        &   $\mu$m        \\ \hline
$Z$ &  0.8817 & 0.836$-$0.929 & 0.093 \\
$Y$ &  1.0305 & 0.979$-$1.081 & 0.102 \\
$J$ &  1.2483 & 1.169$-$1.328 & 0.159 \\
$H$ &  1.6313 & 1.492$-$1.784 & 0.292 \\
$K$ &  2.2010 & 2.029$-$2.380 & 0.351 \\ \hline
\end{tabular}
\end{scriptsize}
\end{table}

\subsection{Plan of paper}  \label{paperplan}

%

This paper begins with a description of the science goals of UKIDSS, and some
illustrations of how the survey design will achieve them. We then describe the
practical implementation of the survey - the tiling and jittering patterns,
exposure times, calibration plan, and so on, in the context of the camera
properties and the UKIRT operating procedures and software. Next we consider
the detailed design of the individual survey components - areas, field
selection, filters and scheduling. We also describe the staging of UKIDSS in a
two year plan and final seven year plan. We then summarise the data processing
arrangements, which as described above are pursued as a formally separate
project, but which of course are crucial to the scientific success of UKIDSS.
Following this we present some example data and simple analysis from the science
verification phase of UKIDSS, and point towards the expected final data quality.
Finally we summarise progress to date, including an analysis of how well
UKIDSS is meeting its design targets, describe the plan for publication of the data, 
and provide links to
more detailed information about UKIDSS, WFCAM, and the science archive.

\section{Science goals}  \label{sectn:survey-goals}

\subsection{General goals}

The primary goal of UKIDSS is to produce IR sky atlases and catalogs as a fundamental
resource of lasting significance analogous to the various Schmidt photographic
sky surveys of the 1970s and onwards (Hambly \et 2001 and references therein), 
and to the SDSS survey of modern
times (York \et 2000). None of our survey components
covers the whole sky, but none the less each component deserves the
term ``atlas'', as the volume surveyed, and the number of objects detected are
comparable to the above optical surveys, and each survey maps out some
significant part of the universe
- the solar neighbourhood, the Milky Way, the local extragalactic universe, the
universe at z=1, and the universe at z=3. Each of the component surveys is many
 times larger than any existing IR survey at comparable depth.
 
%
\begin{table*}
\begin{tabular}{lccccccc}
\bf Survey Name \rm & \bf Passbands \rm & \bf Depth \rm & \bf Area \rm & \bf Seeing \rm & 
\bf Phot \rm  & \bf Astrom \rm & \bf Ellipticity \rm \\ 
 \it Notes \rm  &      &   K mag        &  sq.deg      & FWHM   & RMS  & RMS    &    \\
             &      &                &              & arcsec & mag  & arcsec &      \\
             &      &                &              &        &      &        &      \\
            
Large Area Survey (LAS)  & YJHK & 18.2 & 4028 & $<$1.2 & $<$0.02 & $<$0.1 & $<$0.25 \\
\it Matched to SDSS areas \rm &&&&&&& \\ 
Deep Extragalactic Survey (DXS)  & JK & 20.8 & 35 & $<$1.3 & $<$0.02 & $<$0.1 & $<$0.25 \\
\it Four multiwavelength fields \rm &&&&&&& \\ 
Ultra Deep Survey (UDS)  & JHK & 22.8 & 0.77 & $<$0.8 & $<$0.02 & $<$0.1 & $<$0.25 \\
\it Subaru/XMM Deep Survey Field \rm &&&&&&& \\ 
Galactic Plane Survey (GPS)  & JHK & 18.8 & 1868  & $<$1.0 & $<$0.02 & $<$0.1 & $<$0.25 \\
\it Northern Plane, $\left|b\right|<5^\circ$ \rm &&&&&&& \\ 
Galactic Clusters Survey (GCS)  & ZYJHK & 18.6 & 1067 & $<$1.2 & $<$0.02 & $<$0.1 & $<$0.25 \\
\it Ten large open clusters \rm &&&&&&& \\ 

\end{tabular}


\caption{\label{table:survey-goals}\it General design of each component of UKIDSS. These are the
planned survey goals as originally approved by the UKIRT Board, together with survey quality parameters applied during preparation for the First Data Release.  The final areas and depths will depend on a variety of factors including survey efficiency and UKIRT time allocation.  }

\label{surv-goals}

\end{table*}


The strength of a survey is of course its potential for multiple use over many
 years, but this general aim does not fix the best combination of area, depth,
 and wavelength coverage. In a Euclidean volume, for a given total time, a
 shallow survey always produces a larger sample size than a deep one; but
 specific science goals often require a given depth, for example to detect
 galaxies at a given redshift; and for relatively deep surveys, neither the
 Milky Way nor the universe at large are Euclidean volumes. As we cannot predict
 all future uses of the UKIDSS survey databases, the general idea is to pursue a 
 ``wedding cake'' strategy, dividing the time between a large shallow survey, a
 medium sized fairly deep survey, and a small very deep survey, and including
 targeted observations of the Galactic Plane and nearby clusters. 
 
 The general features of 
 the five planned survey components are described in Table \ref{table:survey-goals}. 
 (More details are given in section \ref{survey-design}  
 and in the data release papers, 
 Dye \et (2006) and Warren \et (2007a,b)).
The intention is to achieve uniformity of depth across each survey.
\footnote{The depths quoted in this table and elsewhere are magnitudes
in the Vega system for a point source which attains a 5$\sigma$ detection within a 2\arcsec aperture.}
Separate limits apply in each band of course. The photometric and astrometric accuracy goals are requirements set by the UKIDSS consortium on the VDFS data processing system. They are maximum absolute errors, and so include any non-uniformity across the survey. Seeing and ellipticity `goals' in the table are a simplified version of the cuts that have been applied in quality control filtering during preparation for the first data release; the actual cuts vary between bands, and the great majority
 of dataframes easily pass these cuts.
 
In the following
 subsections we summarise the scientific goals of each survey component, and
 overall survey quality goals.
 In section \ref{survey-design} we describe the design of each survey component - 
 areas, depths, field locations, filters, implementation strategy - needed to achieve
 those goals. In section \ref{sectn:survey-progress} we summarise the progress so far towards
 the design goals, and the
 actual achieved quality.

\begin{figure}

\includegraphics[width=80mm,angle=0,clip]{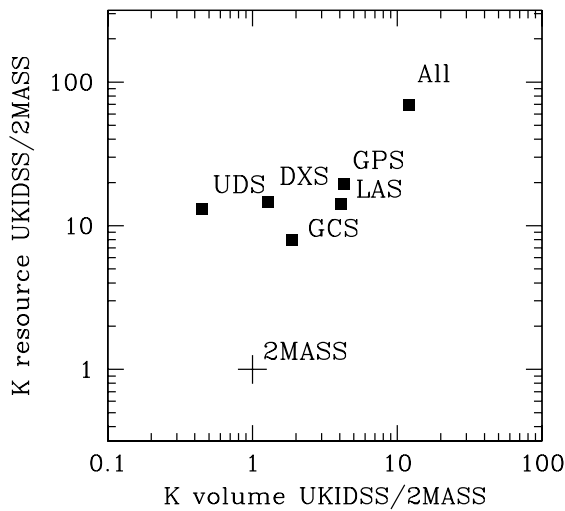}

\caption{\label{fig:udef-goals-general-depth-comp} \it Illustration of the scope 
of the five UKIDSS survey components,
and their sum, 
by comparison with 2MASS. The comparison is made in terms of effective
volume, and effective resource, for the K band,
computed as described in the text}

\end{figure}

The scope of the surveys is illustrated in an interesting way in 
Fig. \ref{fig:udef-goals-general-depth-comp}.  
For sky limited observations in the K-band,
one can show that for a uniform source population in Euclidean space,
the effective volume surveyed is proportional 
to the quantity  Area $\times10^{0.6K}$. 
On the other hand, 
the time to reach depth K is proportional to $10^{0.8K}$. 
So, as one surveys the same area deeper, the effective volume increases
slowly, but the overall information gained (for example on colours of brighter
sources) increases somewhat faster.
For fixed observing conditions, the quantity $10^{0.8K}$ is 
proportional to etendue $\times$ instrument throughput $\times$ time spent, 
which we can think of as overall `effective survey resource'. 
The largest existing multiband near-IR 
survey in terms of both quantities is 2MASS. In 
Fig. \ref{fig:udef-goals-general-depth-comp}
we have 
normalised the computed values for each of the five UKIDSS elements, to 
the 2MASS values using the 2MASS limit of K=15.50 (Skrutskie \et 2006).
 Viewed in this way, each of the 5 
K-band surveys is between 10 and 30 times larger than 2MASS in terms of 
resource, and, except for the UDS, is a few times larger in 
volume. Summed over the whole programme, UKIDSS is 70 times larger 
than 2MASS in terms of resource, and 12 times larger in terms of volume.

In practice, achieving the science goals of UKIDSS will depend not just on
the depths and areas achieved, but also on the survey quality - completeness,
reliability, number of artefacts and so on. These issues are discussed in 
section \ref{sectn:data}, with some preliminary results given in section 
\ref{sectn:survey-progress}.


%

\subsection{Headline science goals}

To develop a plan beyond this very general concept, we encouraged the formation
 of groupings within the consortium to promote distinct survey components and
 science goals. The whole consortium then debated the science goals and designs
 proposed by these groups, looked for overlaps, and made compromises until we
 felt we had a balanced strategy. (The resulting survey component designs are
 presented in Section \ref{survey-design}.) As part of this scientific debate, we also
 agreed {\em headline science goals}, which were used to drive the specific
 designs of the survey components. The most important of
these specific goals are as follows :

\begin{itemize}

\item to find the nearest and faintest sub-stellar objects

\item to discover Pop II brown dwarfs, if they exist

\item to determine the substellar mass function

\item to break the z=7 quasar barrier

\item to determine the epoch of re-ionisation

\item to construct a galaxy catalogue at z=1 as large as the SDSS
catalogue

\item to measure the growth of structure and bias from z=3 to the present
day

\item to determine the epoch of spheroid formation

\item to clarify the relationship between quasars, ULIRGs, and galaxy
formation

\item to map the Milky Way through the dust, to several kpc

\item to increase the number of known Young Stellar Objects by an order of
magnitude, including rare types such as FU Orionis stars

\end{itemize}

\subsection{Goals of the Large Area Survey (LAS)}
\label{sectn:goals-las}

The Large Area Survey (LAS) aims to map as large a fraction of the
Northern Sky as feasible (4000 square degrees) within a few hundred
nights, which when combined
with the  SDSS,  produces an atlas covering almost an order of magnitude
in wavelength. Furthermore a huge number of objects will already have
spectroscopic data from the SDSS project.
The target depths of the basic shallow survey are Y=20.3, J=19.5, H=18.6, K=18.2. We also
plan a second pass in the J-band, with an average epoch separation of 3.5 years, 
and a minimum separation of 2 years, 
to detect proper motions of low
mass objects and thus their kinematic distances. The 
final J depth target
is therefore J=19.9. (The final achieved depth may
of course be different; 
this is discussed in \ref{sectn:survey-progress}.)

The Large Area Survey, when combined with the matching SDSS data,
will produce a catalogue of a half a million galaxies with
colours and spectra, and several million galaxies with photometric redshifts;
will detect thousands of rich clusters
out to z=1; will find ten times more brown dwarfs than
2MASS, will probe to much fainter objects, and can get statistical ages
and  masses from kinematics; and will produce a complete sample of 10,000
bright quasars, including reddened quasars, using the K excess method
(Warren, Hewett and Foltz 2000). 

\begin{figure}

\includegraphics[width=80mm,angle=0]{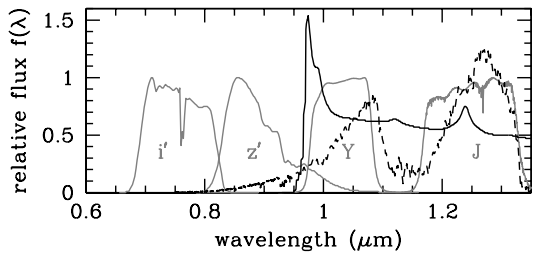}

\caption{\label{fig:udef-goals-las-seds}
\it Plot illustrating the usefulness of the Y band for finding cool 
brown dwarfs and quasars of very-high redshift ($z>6.4$). Filter curves 
are total system throughput (above atmosphere to detector), normalised 
to the peak, for the SDSS i' and z' bands (from Fan et al. 2001), and 
the WFCAM Y and J bands (from Hewett et al. 2006). The dashed curve is 
the spectrum of the T6  brown dwarf SDSS J162414.37+002915.6 (from 
Leggett et al. 2000), and the solid curve is a model spectrum of a 
quasar at z=7. High-redshift quasars and brown dwarfs may be identified 
by the very sharp spectral discontinuity in moving from the optical 
(i', z') to the near-infrared (Y, J), while the quasars may be 
distinguished from the brown dwarfs, because they are
somewhat bluer in Y-J 
colour.}

\end{figure}

We are particularly driven however by three especially exciting
prospects. 
(i) A search for the the nearest and smallest objects in the
solar neighbourhood.The LAS is deep enough to detect brown dwarfs and 
young ($<$ 5 Myr) free floating planets with as little as 5 Jupiter masses out to 
distances of tens of parsecs. 
The LAS should find brown dwarfs even 
cooler than T dwarfs, $T_e<$700K, a new spectral class tentatively named Y 
dwarfs (Leggett \et 2005). (ii) The combination of IR and optical colours, and
large expected proper
motions, will allow the LAS to find halo brown dwarfs if they exist, testing
the universality of star formation processes, and the formation history of
the Milky Way. (Calculations based on 
Burrows \et (2003) indicate that halo brown dwarfs will have an 
absolute Y magnitude $>$16 and so could be found to distances of 60pc.) 
(iii) We hope to find quasars at $z=7$ and to detect the
epoch of re-ionisation. SDSS
have found z=5-6 quasars by  ``i$^\prime$ drop-out''. Beyond z=6
quasars become rapidly redder, indistinguishable from brown dwarfs in
standard colours, and too faint to be in the SDSS $z^\prime$ survey. We
therefore intend to undertake a survey in the new Y filter to match our JHK
survey. Extrapolating the evolution function of Fan \et\ 2001 to predict
quasar numbers detected in both Y and J to 10 $\sigma$, 
the LAS should find
10 quasars in the range z=6-7 and 4 in the range z=7-8. Figure \ref{fig:udef-goals-las-seds}
illustrates how the UKIDSS filter set can distinguish cleanly between very cool
brown dwarfs and very high redshift quasars.


%

\subsection{Goals of the Galactic Plane Survey (GPS)}

The Galactic Plane Survey aims to map half of the Milky Way to within 
a latitude of $\pm 5^{\circ}$. Given the declination constraints of UKIRT, 
we can survey l=15$^{\circ}$--107$^{\circ}$ and 
l=141$^{\circ}$--230$^{\circ}$. Owing to interest in  recent results from
multi-waveband observations of the Galactic Centre region 
(eg. Wang \et 2002; Hasegawa \et 1998) 
the survey region has been extended south
to include the l=-2$^{\circ}$ to 15$^{\circ}$ region in a narrow strip at $b=\pm 2^{\circ}$.
The target depths are J=19.9, H=19.0, K=18.8, with the K depth made up
of three separate epochs with a depth of K=18.2 each time.\footnote{These 
depths refer to uncrowded regions well away from the Galactic Centre 
and a few degrees out of the plane. In crowded regions the survey will be less deep, 
due to added background noise from unresolved stars (see section 6.2)} 
This is deep enough to probe the IMF down to 
M$\sim 0.05$~M$_{\odot}$ in star formation regions within 2~kpc of
the sun, to detect stars below the main sequence turn off in the galactic 
bulge, and to detect luminous objects such as OB stars and post-AGB stars
across the whole galaxy. The K band repeats are for both extra depth and
to detect highly variable objects
and locate nearby objects through their proper motions. In addition we
will make a three epoch narrow band H$_2$ survey in a 300 square degree 
area of the Taurus-Auriga-Perseus molecular cloud complex (with JHK data also).
This survey area closely follows the region of molecular emission detected
by Ungerechts \& Thaddeus (1987).

Like the high latitude LAS, the GPS has its prime importance as a fundamental
resource for future astronomy. The survey depths are close to being confusion
limited, so this survey is unlikely to be superseded until a high resolution
wide angle camera is placed in space. We expect to detect several times $10^{9}$ sources
in total. However, there are a number of immediately expected results,
which will be achieved in combination with data from multiwaveband 
galactic surveys from many facilities. There will be particular benefit
from surveys planned or in progress with the Isaac Newton Telescope (optical),
SPITZER space telescope 
(infrared), CHANDRA and XMM-Newton (X-ray), the VLA (radio, especially the 
5~GHz CORNISH survey), HERSCHEL and SCUBA-2 (submm) and AKARI (far infared).
The following list illustrates some of the expected results.
(1) An increase in the number of known Young Stellar Objects (YSOs)
by an order of magnitude and measurement of the duration of the YSO phase as
a function of mass and environment. (2) Star formation regions will be mapped
throughout the Milky Way, measuring the environmental dependence of the
IMF to low masses and estimating the overall star formation rate of the galaxy.
(3) Rare or brief duration variables will be found in significant numbers,
aiding the study of phenomena such as 
FU Orionis variables, Luminous Blue Variables and unstable post-AGB stars undergoing 
thermonuclear pulsations. (4) Thousands of evolved objects such as 
protoplanetary nebulae and planetary nebulae will be found, a huge increase over 
previous samples. This will be achieved by selecting candidates from far 
IR surveys such as the Akari (ASTRO-F) survey. New detections are expected
to be commonest for small protoplanetary nebulae and very large low surface
brightness planetary nebulae, which would respectively be unresolved or undetected 
by 2MASS.
(5) Many stellar populations will be mapped to large distances through the 
Milky Way extinction, measuring the scale height versus stellar type and 
mapping poorly measured regions of the arms and warp. (6) The IR counterparts 
of 
hundreds of X-ray binaries, thousands
of CVs, and thousands of coronally active stars will be identified
and 
source lists provided for regions yet to be mapped by X-ray satellites.

\subsection{Goals of the Galactic Clusters Survey (GCS)}

The Galactic Clusters Survey (GCS) aims to survey ten large open star
clusters and star formation associations, covering a total of 1067 sq.deg.
using the standard single pass depth (see section \ref{wfcam-obs})
plus a second pass in K for proper
motions, giving a depth of Z=20.4, Y=20.3, J=19.5, H=18.6, K=18.6. The 
targets are all relatively nearby, are at intermediate to low Galactic 
latitudes and are several degrees across.

\begin{figure}

\includegraphics[width=60mm,angle=270]{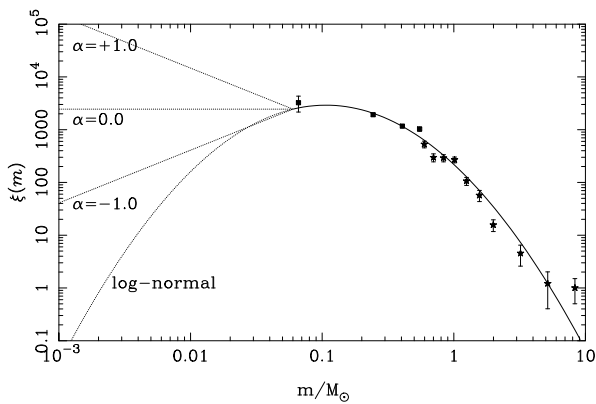}

\caption{\label{fig-goals-gcs-imf}
\it Various extrapolations (dotted lines) of the Pleiades mass
function (after Hambly et al. 1999) illustrating uncertainties in
the behaviour of the Mass Function (MF) in the brown dwarf (BD) regime. (Vertical axis
is cluster stars per unit mass.) The most recent
surveys have probed the mass range 0.01 to 0.1 solar masses, using
a variety of techniques, and have produced a range of different
forms of the MF using heterogeneous datasets with varying degrees
of completeness. The GCS aims to obtain maximal completeness in
ten targets to settle the questions as to the form and universality
(or otherwise) of the MF in the BD regime.
}

\end{figure}

The GCS is the most targeted of our surveys, being aimed at the crucial
question of the sub-stellar initial mass function (IMF).
Our current knowledge of the IMF is illustrated in Fig. \ref{fig-goals-gcs-imf}, 
by comparing measurements of the Pleiades with various functional forms.
(Note that determining whether the Pleiades is typical or not is part
of the goals of the survey).
The stellar IMF is well determined down to the brown dwarf
boundary but is much less well known below, and it is not known whether the
IMF as a whole is universal or not (the current state of research into
ultra low--mass star formation is described in Mart\'{i}n \& 
Magazz\`{u}~2007). The mass limit reached varies somewhat
from cluster to cluster, but is typically around $M_L \sim 30 M_J$, where $M_J$ 
is the mass of Jupiter. The
number of objects expected to be detected in the range $M_L$ -- $M_L +
10M_J$ ranges from 100 to 3000 for the range of possible mass function
models, showing how well we will constrain the IMF compared to current
knowledge.


%

To find extreme objects --- the very nearest examples, the lowest mass 
objects --- the Large Area Survey is better. But to measure the 
substellar IMF, one wants to target the 30 -- 100 $M_J$
region, and to obtain masses one needs both a distance and an age, for
which mapping clusters is ideal. This approach has of course already been
started (e.g.\ Moraux \et ~2007 and references therein).
Our survey improves on current studies not by going deeper but by
collecting much larger numbers, and examining objects formed in
environments having a range of ages and metallicities, to examine the 
question of universality.

\subsection{Goals of the Deep Extragalactic Survey (DXS)}

\begin{figure}

\includegraphics[width=75mm,angle=0]{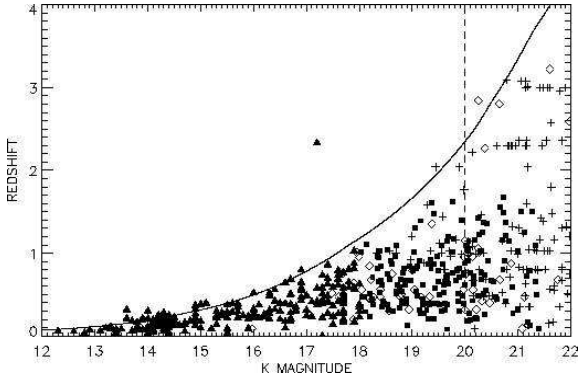}

\caption{\label{fig:udef-goals-dxs-kz} \it The redshift distribution of a K selected galaxy sample from the Hawaii Deep Fields. This is an updated version of the figure in Songaila et al (1994), kindly provided by L.Cowie and collaborators. The solid symbols show spectroscopic redshifts from the Hawaii Deep Fields, which have been completely observed to K=20 (dashed line) though only identified objects are shown. The open diamonds show the spectroscopically identified objects in the Hubble Deep Field, while the crosses show all the remaining objects at their photometric redshift. The solid line shows the K magnitude of a 2L* unevolving Sb galaxy.}

\end{figure}

The Deep Extragalactic Survey (DXS) aims to map 35 sq.deg. of 
sky to a 5-$\sigma$ point-source sensitivity of J=22.3 and K=20.8 
in four carefully selected,
multi-wavelength survey areas. Central regions of each field
will also be mapped to H=21.8.
The primary aim of the survey
is to produce a photometric galaxy sample at a redshift of 1--2, within
a volume comparable to that of the SDSS, selected in the same
passband (rest frame optical). Figure \ref{fig:udef-goals-dxs-kz} shows 
measured K magnitude versus redshift for galaxies in the Hawaii Deep Fields
(L.Cowie, private communication). This shows that to achieve a sample 
such that the median redshift 
is $z\sim 1$ requires measuring galaxies with $K\sim 20$ and so
going to a point source depth of K$\sim$21.  
Such a sample will allow a direct
test of the evolution of the galaxy population and determine how 
galaxies of different types (passive, star-forming, AGN)
trace large scale structure (their bias). Each of these
properties can be predicted from cosmological simulations so
the DXS will set tight constraints on these models in 
volumes less susceptible to cosmic variance than previous,
narrow-angle surveys at this redshift. The
sample will also enable the selection of clusters of galaxies
in this redshift range, where cosmological models predict
numbers to be sensitive to the total mass density of the
Universe, $\Omega_0$. 

The number of deep, multi-wavelength survey fields has increased
dramatically in the past 5 years with the up-grade of
existing facilites (e.g. Megacam on CFHT and VIMOS on the VLT) and new satellite
missions (e.g. {\it Spitzer}, {\it GALEX}, {\it XMM-Newton} and
{\it Chandra}). Each of these facilities has current surveys
of 2--40 sq.deg. of contiguous area to levels where many
of the counterparts are intrinsically faint in the optical (R$>$24)
due to a combination of redshift and/or intrinsic dust obscuration,
but are relatively red (R-K$>$4). Therefore deep NIR imaging (K$\sim$21) over
tens of square degrees is required to fully characterise these 
dusty and/or distant objects. Looking ahead to the end of
this decade and the completion of UKIDSS, there will be 
many more complementary surveys on these scales
in other wavebands such as the far-infrared ({\it Herschel}), sub-mm (SCUBA-2),
radio (EVLA and eMerlin) and Sunyaev-Zel'dovich (SZ telescope and AMI).
The legacy potential of the DXS was a key driver for the science
case and the field selection (see section 4.4).


%

\subsection{Goals of the Ultra Deep Survey (UDS)}

The Ultra Deep Survey (UDS) aims to map $0.77$ sq. degrees to a 5$\sigma$
point-source sensitivity of J=24.8, H=23.8, K=22.8.  Such depths are required to
reach typical $L^*$ galaxies at $z=3$. Covering an area one hundred times
larger than any previous survey to these depths, this will provide the
first large-volume map of the high-redshift Universe ($30\times 30$Mpc by
2 Gpc deep at $2<z<4$).

Deep near-infrared surveys are crucial for obtaining a more complete
census of the Universe at these epochs. In particular, galaxies which are
reddened by dust or those which appear red due to old stellar populations
may be completely missed by standard optical surveys. From the UDS we
anticipate over 10,000 galaxies at $z>2$, allowing detailed studies of the
luminosity functions, clustering and multi-wavelength SEDs over a large,
representative volume. A major goal, together with the DXS and local
surveys, is to measure clustering as a function of stellar mass and cosmic
time, which will provide very powerful tests of models for biased galaxy
formation and the growth of structure.

The UDS is also designed to address one of the major unsolved problems in
modern astronomy, which is to understand when the massive elliptical
galaxies are formed. A key test will be to determine the co-moving number
density of the most massive galaxies at various epochs, particularly at
$z>2$. This requires a combination of both depth and area which has
previously been impossible to achieve.  If the density of massive galaxies
(more massive than local $L^*$ ellipticals) is similar to that of today,
we should see $\sim 1000$ per square degree. Current semi-analytic models
predict an order of magnitude fewer. Our goal is to directly measure the
build-up of this population over cosmic time.

The survey field chosen for the UDS is the Subaru/XMM Deep Field, which
has a wide range of multiwavelengh data available, including deep radio
observations from the VLA, submm mapping from SCUBA, mid-IR photometry
from Spitzer, deep optical imaging from Subaru Suprimecam and deep X-ray
observations from XMM-Newton.  When combined, these will enable detailed
studies of the relationship between black hole activity, dust-dominated
ULIRGs and IR-selected massive galaxies using an unprecedented
high-redshift sample.

\section{Implementation with the UKIRT Wide Field Camera}  \label{WFCAM-imp}

\subsection{General characteristics of telescope and camera}

UKIDSS is implemented using the Wide Field Camera (WFCAM)  on the United
Kingdom Infrared Telescope (UKIRT), which is operated by the Joint Astronomy
Centre (JAC), an establishment of the UK's Particle Physics and Astronomy
Research Council (PPARC).  General technical details for UKIRT are given on the
JAC website\footnote{http://www.jach.hawaii.edu/UKIRT/}.  It is an infra-red dedicated
3.8m telescope operating at the summit of Mauna Kea in Hawaii. Of particular
importance is a tip-tilt secondary, which primarily removes dome and
windshake effects on seeing, delivering close to free-atmosphere seeing (half
arcsecond on many nights) across the whole WFCAM field of view. With the advent of
UKIDSS, UKIRT now operates in part as a survey telescope and in part 
as an open access telescope offering time through periodic peer-reviewed competition.
At the time of writing, WFCAM is
scheduled for 60\% of UKIRT time, 220 nights per year. After removal
of engineering time, and time allocated to the University of Hawaii, and Japan,
an average of 167 nights per year is left, 80\% of which, 134 nights per year, is
devoted to UKIDSS, and the remaining time to other peer-reviewed programmes. 
(The latter are selected by the UK PATT system, open to world wide proposals).

\begin{figure}

\includegraphics[width=70mm,angle=0]{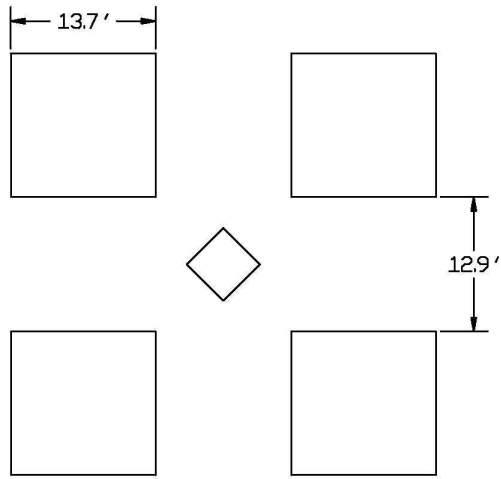}

\caption{\label{fig:udef-focal-plane} \it
The WFCAM focal plane. The spacing between detectors is 94\% of the
width of each detector. A sequence of four pointings therefore
produces complete coverage for one ``tile'', plus a small overlap
region. The central diamond is the autoguider CCD. For details see
Casali \et (2007).}

\end{figure}

WFCAM is described in detail by Casali \et (2007).  Here we summarise some
key characteristics. WFCAM has an unusual design, with an array of IR detectors
inside a long tube mounted above Cassegrain focus. The forward-Cassegrain
Schmidt-like camera design makes possible a very wide field of view (40 arcmin)
on a telescope not originally designed for this purpose. The camera  has four
2048$\times$2048 Rockwell Hawaii-II PACE arrays. The arrays have a projected
pixel size of $0.4\arcsec$, which gives an instantaneous exposed field of view
of 0.207$\,$deg$^2$ per exposure. The arrays are spaced by 0.94 detector widths.
The focal plane coverage is illustrated in Figure \ref{fig:udef-focal-plane}.

%

The WFCAM filters, the optical performance, and the detector efficiency
are presented in Casali \et (2007). The photometric system is described in Hewett \et (2006).

\subsection{Observing with WFCAM} \label{wfcam-obs}

WFCAM makes short exposures on the sky, typically 5-10 seconds.
Coverage of the sky is then built
up in several stages - by small jittering patterns around a fixed telescope pointing
position; by macrostepping to make a filled in ``tile''; and by accumulating sets
of such tiles to gradually cover the sky, or revisiting tiles to build up depth.

{\em Integration overheads.} Each WFCAM array  has a separate SDSU
controller, reading out 32 channels (8 segments in each of 4 quadrants). 
The total readout time is 0.7 seconds, and the reset process is almost instantaneous,
but normal procedure involves two dummy readouts to stabilise the array, which
take 0.3 seconds. The overall {\em exposure overhead} is therefore 1.0 seconds. 
Multiple {\em exposures}
can be made
at the same pointing position, co-adding into the same data file. The sum of
these exposures is an {\em integration}, which results in a set of 
four data files which in offline processing are later linked together as a 
{\em multi-frame}. For some users of WFCAM there may be several exposures
per integration, but UKIDSS procedures always have only one exposure 
per integration. In addition each integration has a {\em data acquisition overhead} of 2-3 seconds. 
It is hoped
that hardware and software improvements will improve this in due course, but for the moment there is then a total {\em integration overhead} of 3-4 seconds.
Exposures of 5-10 seconds are therefore long
enough to be reasonably efficient ($\sim$ 55-77\%) and for sky background noise to be larger than the
readout noise. The main exception is in the Y and Z bands, where exposures of 20
seconds are needed for the sky noise to exceed readout noise. Exposures longer
than this are not normally used, as an increasing number of stars in the field
are saturated.

{\em Jittering and micro-stepping.} Small accurate telescope offset patterns 
relative to a fixed base position are used to improve WFCAM data. 
The first method is to
use a jitter sequence with offsets equal to whole numbers of pixels, resulting
in frames which can be co-added. The aim of such a {\em jitter pattern} is to
minimise the effects of bad pixels and other flat-fielding complications. A
variety of jitter patterns can be used. 
The second kind of pattern is {\em
microstepping}, which uses offsets with non-integer numbers of pixels. 
In 2$\times$2
microstepping, offsets by N+1/2 pixels are used. The data are then
interlaced (i.e. keeping the pixels independent) into a grid of
pixel-spacing 0.2\arcsec, producing an image of size
4096$\times$4096 pixels for each array. In 3$\times$3 microstepping,
offsets by N+1/3 and N+2/3 pixels are used. The aim of such a microstepping
pattern is to improve image sampling - the WFCAM pixel size of 
0.4\arcsec\ is adequate for moderate seeing (i.e. $>$0.8\arcsec), but undersamples the
expected seeing a significant fraction of the time.
For sampling and/or cosmetic reasons, all UKIDSS surveys use
at least 2 offset positions, and most use 4 offset positions. 
The ``standard shallow observation'' has a 
total integration time of  40 seconds, with individual exposures chosen to add up to this total. 
The data from offset sequences are normally interleaved and/or stacked offline to make a single multi-frame data file, which is the basic unit of the archived data.
Small offsets (less than 10\arcsec) are intended to have minimal overhead. They
are applied by the tip-tilt system, with a smooth restoration to the optical axis
during observing. Larger offsets are applied by the telescope mount, and observing
commences after the system has stabilised.

{\em Tiling.} The WFCAM arrays are spaced by 0.94 detector
widths. The sky could potentially be covered in a variety of mosaic patterns,
but the typical procedure would be to expose in a pattern of four macro-steps to
make a complete filled-in ``tile''.  Allowing for overlaps with
adjacent tiles, the width of a single tile is then 3.88 detector widths, i.e.
0.883$^\circ$, giving a solid angle of 0.78 sq.degs.  The time between macrostep
integrations (slew, stabilise, guide star lock) is $\sim$ 15 seconds. For
shallow surveys, where each pointing has typically 4 offset positions each with
a 10 second exposure and a 3.5 second overhead, this means that a tile with 4
pointings spends 160 seconds exposing out of an elapsed time of 276 seconds,
making a total {\em observing efficiency} of 58\%. For deeper surveys, where
many integrations are made between telescope slews, the macro-step overhead is
negligible, and the efficiency tends towards $\sim$75\%.

{\em Schedule Blocks and Survey Definition.} UKIRT operates  an automated
flexible queuing system. A precise sequence of
exposures, offset patterns, and filters at each of a list of pointing positions,
which can be thought of as grouped into ``tiles'' as appropriate, is specified in advance. 
These Observations are grouped into ``Minimum Schedulable Blocks (MSBs)'', occupying roughly 20-60 minutes.
The MSBs also
contain constraints that determine whether they can be observed - required
seeing, sky brightness, etc.  
Calibration observations - twlight flats, standard
stars, etc - are entered as independent MSBs. 
The MSBs are entered into a database which is queried during observing
to generate a priority ranked list of MSBs for which the current weather
and observing conditions are suitable. The observer selects an MSB from
this list (normally the highest priority MSB) and sends it onto an
execution queue to be observed. If the MSB is sucessfully completed, it
is marked as such in the database and will not be listed in future query
results. (Occasionally MSBs are partially completed or fail the later Quality Control (QC) 
process, 
and so can be repeated).
For UKIDSS, a Survey Definition Tool (SDT)
is used to design the list of pointing positions and associated guide stars 
for each survey, which are then
grouped into MSBs, and likewise into smaller ``projects'' which help planning and
monitoring of survey completion.

Actual survey data collection rate in practice is considered in section 
\ref{sectn:survey-progress}. Implementation details, such as seeing and sky brightness
constraints,  are given in the data 
release papers - Dye \et\ (2006) and Warren \et (2007a,b).

\subsection{Survey calibration}

The UKIDSS data are calibrated to magnitudes in the Vega system.
The WFCAM photometric system - filter response curves, and synthetic colours
for a variety of objects - is described in Hewett \et (2006). (Approximate passbands 
are given in section \ref{techpapers}). 
Calibration on the sky is achieved using observations of 2MASS stars
within each field, which allows us to derive photometric calibration
even during non-photometric conditions, including colour equations 
for transformation from the 2MASS system to the WFCAM system. 
For the shallow surveys, the Quality Control (QC) 
process includes frames whose zeropoint is within
0.2 magnitudes of the modal value. (The vast majority of accepted 
frames are much better than this.) For the stacked surveys, where there
is an additional self-calibration check, we accept frames within 0.3 magnitudes. 
It is possible that within such occasional frames with modest extinction there will
be some noticeable spatial opacity gradient. This possibility is not included
in the current pipeline, but we intend to examine it in later re-processings.

There are
plenty of unsaturated 2MASS stars in every data frame - in the range 60-1000, 
dependendent on Galactic Latitude.
Furthermore the 2MASS global
calibration is accurate to better than 2\% across the entire sky (Nikolaev \et 2000). 
The procedure
is to cross-match objects detected by the pipeline with 2MASS unsaturated 
sources that have $\sigma_{JHK}(2MASS) \le 0.1$, and to transform the photometry
of these stars into the WFCAM ZYJHK system using empirically
derived colour terms. After correcting counts for the known radial variation in pixel scale, 
the average of these stars gives a global per-frame zeropoint. 
Tests against observations of UKIRT faint standards (Hawarden \et 2001) indicates that this procedure gives us a JHK photometric system accurate to 2\%, which is the survey design requirement. (At the time of writing, the quality of the Z-, Y- and
narrowband-filter calibration has not yet been quantified.) Calibration
from 2MASS stars therefore seems justified.
However, we have also made freqent observations of UKIRT faint standards which provides a backup calibration, and an independent
method of deriving colour equations. The calibration procedure, and the final colour equations between various systems, will be presented in full in Hodgkin \et (in preparation).

\section{Survey Design}  \label{survey-design}

The design of the UKIDSS survey components was driven by a mixture  of the
legacy ambition, practical limitations, and specific science goals. The total
size of the project was chosen by a decision to continue long enough 
to achieve a product of international significance and lasting value. The
total time available was driven by UKIRT/WFCAM scheduling
constraints. We thus arrived at a {\em seven year plan} totalling approximately
a thousand nights. Seven years is however a long time to wait for science
results; we therefore also designed an initial {\em two year plan} that would
produce a self contained product and valuable science.

\begin{figure*}

\includegraphics[width=80mm,angle=270]{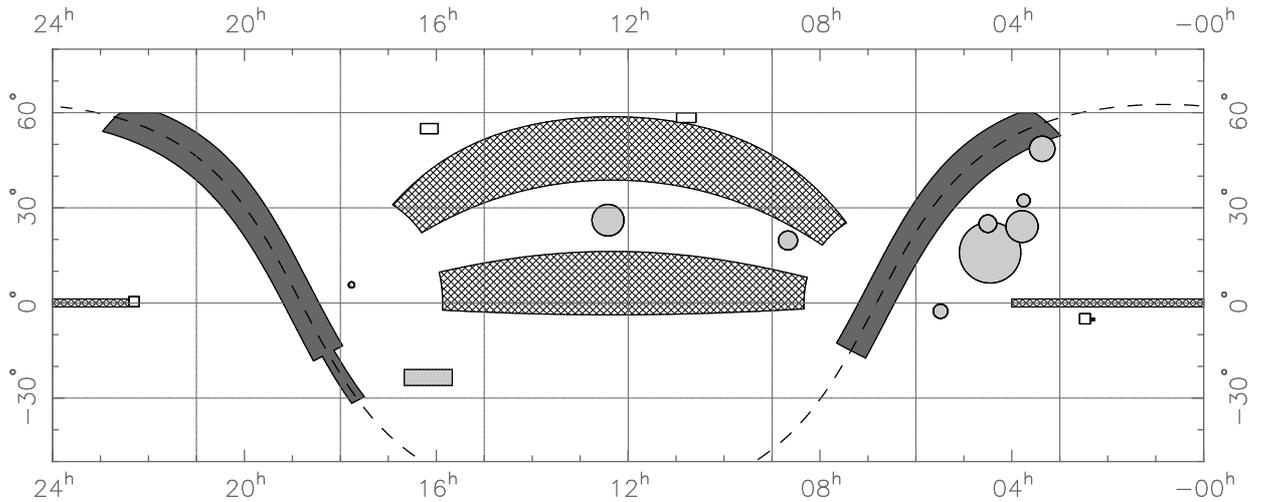}

\caption{\label{fig:udef-design-areas}
\it Location on the sky of the fields comprising the various survey  components.
Cross-hatch : Large Area Survey. Dark Grey : Galactic Plane Survey. Light Grey : Galactic
Clusters Survey. Open rectangles : Deep Extragalactic Survey. Note that the Ultra Deep Survey
lies just to the west of the DXS field at 02h18m - 05$^\circ$10'
The dashed line marks the Galactic plane.
Note that UKIRT lies at latitude
+20$^\circ$. }

\end{figure*}

Table \ref{table:survey-summary} summarises the design parameters of each of the 
five UKIDSS survey components. Figure \ref{fig:udef-design-areas} shows the location of the survey fields on the sky. The
logic behind this design, and some more detail about how the surveys are
implemented, is described below for each survey component in turn. The implementation details were revised after the
first observing block (May-June 2005). We outline the current scheme,
with the expectation that it is unlikely to change significantly. More
complete details will be provided in each paper accompanying milestone
data releases (Dye \et (2006), for the EDR; Warren \et (2007a,b) for DR1 and DR2). 
Note that the survey parameters shown here are the survey goals, 
as approved during the proposal stage, and adjusted to a minor
degree during survey planning. Likewise the times 
quoted are the original proposal estimates. The actual time to reach
these design parameters, or the likely final adjusted survey parameters,
depend on both time allocation by the UKIRT Board and the real survey progress rate,
which is discussed in section \ref{sectn:survey-progress}.  

%
\begin{table*}
\begin{tabular}{lcccccc}
\bf Survey \rm & \bf Area \rm & \bf Filters \rm & \bf Limit \rm
 & $t_{int}$ \rm & \bf $t_{tot}$ \rm & \bf Nights \rm \\

    &      &             &      &      &      &        \\

LAS & 4028 & Y           & 20.3 & 40s  & 367h & 262 nts \\
    & 4028 & J$\times 2$ & 19.9 & 80s  & 734h &         \\
    & 4028 & H           & 18.6 & 40s  & 367h &         \\
    & 4028 & K           & 18.2 & 40s  & 367h &         \\
    &      &             &      &      &      &         \\

GPS & 1868 & J           & 19.9 & 80s  & 286h & 186 nts \\
    & 1868 & H           & 19.0 & 80s  & 286h &         \\
    & 1868 & K$\times$3  & 18.8 & 120s & 495h &         \\
    & 300  & H$_2\times$ 3 & --- & 450s & 237h &        \\
    &      &             &      &      &      &         \\

GCS & 1067  & Z           & 20.4 & 40s  &  86h &  74 nts \\
     & 1067  & Y           & 20.3 & 40s  &  86h &         \\
     & 1067  & J           & 19.5 & 40s  &  86h &         \\
     & 1067  & H           & 18.6 & 40s  &  86h &         \\
     & 1067  & K$\times 2$ & 18.6 & 80s  & 172h &         \\
     &      &             &      &      &      &         \\

DXS & 35   & J           & 22.3 & 2.1h & 415h & 118 nts \\
     & 5  & H           & 21.8 & 4.4h & 124h           \\
     & 35  & K           & 20.8 & 1.5h & 287h           \\
    &      &             &      &      &      &         \\

UDS & 0.77 & J           & 24.8 & 209h & 983h & 296 nts \\
    & 0.77 & H           & 23.8 & 174h & 818h &         \\
    & 0.77 & K           & 22.8 &  58h & 271h &         \\
    &      &             &      &      &      &         \\

\bf TOTAL  & &           &      &      &      & 936 nts \\

\end{tabular}

\caption{\label{table:survey-summary}\it Summary of design goals for
each survey. 
Note that the time estimates quoted are based on the efficiencies etc assumed
at the time of proposal. The real time required to achieve these goals
will be somewhat longer,
and is being assessed at the end of the two year plan.
{\bf\rm Notes :}
(i) Area is in square degrees. (ii) ``J$\times$2'' implies that two passes of the
whole area are made in that filter. (iii) ``Limit'' is the Vega magnitude
of a point source predicted to be detected at 5$\sigma$
within a 2\arcsec aperture. (iv) $t_{int}$ is the
accumulated integration time at each sky position. (v) $t_{tot}$ is
the number of hours required on-sky to complete the survey, allowing for the
estimated exposure efficiency and mosaic efficiency. These efficiency factors
are different for each survey.
(vi) ``Nights'' is the estimated number
of nights required, allowing for calibration and average fraction of useable
UKIRT time (assumed 70\% clear). }

\label{surv_summ}

\end{table*}


%

%

\subsection{Microstepping strategy}

Microstepping improves the sampling, but has the disadvantage 
of making overheads worse. With $2\times2$ microstepping, repeated after an
offset, the exposure time in the shallow surveys (mostly 40s total
integration) is 5s. Without microstepping longer exposure times are
possible, 10s or 20s, which reduces the overheads
significantly. Experimentation
with WFCAM data indicates that microstepping has little advantage for
photometric accuracy, but improves the astrometric accuracy. In each
of the shallow surveys, repeat passes are therefore made in a particular
`astrometric' band, for measuring proper motions; J for the LAS, and K
for the GCS and GPS. As a compromise between increased overheads and
better astrometry, in the current scheme for the high latitude shallow surveys
(LAS and GCS), the
astrometric band is $2\times2$ microstepped, while the other bands are
not microstepped. For the GPS, $2\times2$ microstepping is employed
because of the importance of object separation in crowded fields.
(Of course $3\times3$ would give even better PSF sampling, but this only
makes a significant difference in the small percentage of cases where seeing
is better than 0.3\arcsec.) For the deep surveys, the total integration times in
a field are much larger, allowing the use of longer exposures while
microstepping, so overheads are not an issue. In the DXS $2\times2$
microstepping is used, and in the UDS $3\times3$ microstepping is
used, in all bands.

\subsection{Design of the Large Area Survey (LAS)}

The main science goals of the LAS require as large a volume  as possible, with
increased area being a more efficient use of time than increased depth. However
the survey rate is constrained by our requirement for multiple pass bands, by
the need for exposures long enough to avoid inefficient observing, and by the
need for jittering in order to improve cosmetic quality and/or spatial sampling.

The detailed implementation also depends on the weather constraints used.
Because LAS uses a large fraction of the UKIDSS time, it would obviously not
demand the best seeing. Colours are very important, and some objects are variable,
which argues for doing all four bands when each sky position is visited; on the
other hand, the Y and J observations require darker sky than the H and K
observations.  MSBs were then grouped so that H and K would be done together,
and Y and J done separately. However, the queue is monitored and adjusted to try
to make sure the YJ and HK observations are not too far apart. 
Our current plan is to prioritise uniformity and 
achieve the limits given in section \ref{sectn:survey-goals}. 
Thus in poorer 
seeing conditions the integration times are increased to compensate.

Field selection for the LAS was designed to have a good spread in RA,
to have a reasonable amount of sky coverage at lower declinations, for
follow--up on ESO telescopes, to keep below the UKIRT Declination limit
($+60^\circ$), all while lying with the SDSS footprint.  There are
three sub-areas, shown in Fig \ref{fig:udef-design-areas}. The
detailed field co-ordinates are defined on the UKIDSS web pages.

(i){\em The LAS equatorial block : 1908 sq.deg}. This includes most of SDSS stripes 9 to
16. 

(ii) {\em The LAS northern block : 1908 sq.deg} This includes most of SDSS
stripes 26 to 33. 

(iii) The {\em LAS southern stripe : 212 sq.deg.}  
This is a section
of SDSS stripe 82, extending over -25$^\circ<$RA$<$+60$^\circ$,
-1.25$^\circ<$Dec$<$+1.25$^\circ$. Stripe 82 has been repeatedly
scanned by SDSS, and this is the region of highest quality.

\subsection{Design of the Galactic Plane Survey (GPS)}

The GPS aims to map as much of the Galactic Plane as  possible to a latitude of
$\pm 5^\circ$. The Galactic Latitude limit is chosen to match other surveys, for
example the MSX survey (Egan and Price 1996). The survey area is then largely dictated by the
UKIRT Declination limit to the North, and by the latitude of UKIRT to the South.
Within a reasonable length of time, we can then afford to go roughly a factor of
two deeper than the ``standard shallow observation'' defined in section \ref{wfcam-obs}
This depth is good enough to see all of the IMF to the H--burning limit 
in quite distant clusters, to
see AGB stars all the way through the Galaxy, and to see ordinary G-M stars to
several kpc. The tradeoff to consider is then between depth, colours, and repeat
coverage. At least three bands, and preferably, four are needed, in order to
estimate both spectral type and extinction. 
However, extinction is large enough in much of the Plane that
Y band observations are impractical. For regions of low extinction
the optical IPHAS survey at r',i' and Halpha (see http://www.iphas.org)
will provide sufficient additional colours to determine the
average extinction as a function of distance in each field
using reddening independent colour indices. Other statistical
methods to measure extinction using the JHK colours alone can also
be employed - see Lopez-Corredoira \et (2002).
Measurement of both variability and proper motions is a
goal of the GPS. As a bare minimum to achieve this, we plan three epochs in one
band spread across $\sim$ 5 years. We choose the K-band to make these repeats,
because, given extinction, this is the sensible band in which to build up depth
- it is the K-band that allows us to see clean through the Galaxy in most directions. 
In summary then, we plan an initial pass at JHK, with integrations of 80s,
80s and 40s in the 3 bands respectively, followed by two
further passes at K with 40s integrations at intervals of
at least 2 years for any survey tile.

One  of the goals of GPS is the discovery and study of Young Stellar Objects,
and in particular molecular outflows. We therefore plan in addition to the above
a survey of a single large star forming region in a narrow band H$_2$ filter. (For details
of this filter, see Casali \et 2007).

The  area to be mapped is shown in Fig. \ref{fig:udef-design-areas}. 
The main area is defined by the
Galactic latitude range $b=\pm 5^\circ$, Dec$<$60$^\circ$, and Dec $>$-15$^\circ$.
These constraints define two sections of Galactic longitude, which are
$15^\circ<l<107^\circ$, and $142^\circ<l<230^\circ$. In addition we will map a
narrow extension through the Galactic centre, within $b=\pm2^\circ$, covering
Galactic latitudes $-2^\circ<l<15^\circ$. The Galactic bulge will also be
explored by surveying a thin stripe extending upwards in latitude from the
Galactic centre. Finally, the molecular hydrogen survey maps the
Taurus-Auriga-Perseus complex.

Normal procedure is to do all bands in one visit, as colours are important and
many objects are variable.

\subsection{Design of the Galactic Clusters Survey (GCS)}

The  design of the GCS is relatively simple. It needs to target several separate
clusters, in order to examine the substellar
IMF over a range of ages and metallicities.
The depth requirement is set by the need to detect objects in the 30-100 M$_J$
range. Assuming the standard ``shallow survey'' depth (see section \ref{wfcam-obs}), 
this means that clusters
have to be fairly close and/or young, i.e. within a few hundred parsecs and/or
less than a few hundred million years old. There are relatively few
such objects, and they are several degrees across. A natural strategy therefore
emerges using the standard shallow depth and surveying ten nearby clusters,
covering 1067 sq.\ deg.\ in total. To distinguish cluster members, all five
passbands are needed (ZYJHK), plus a measurement of a proper motion using a second
pass in the K band. Full colour information provides cluster sequence
discrimination in multi--colour space, reddening estimates using shorter versus
longer wavelength colour indices, breaking the degeneracy between reddening due to
instellar extinction and that due to the presence of circumstellar disks.

The  strategy is to cover the majority 1067 sq.\ degs in a single pass in K, to
provide the proper motion baseline, within the initial two year plan.
Table \ref{table:gcs-clus} lists the parameters of the chosen clusters.

%

\begin{table*}
\begin{tabular}{llllllll}
\bf Priority/ \rm & \bf Type \rm & \bf RA \rm & \bf Dec \rm & \bf Area \rm & \bf Age \rm & \bf Minimum \rm \\

 \bf Name \rm &  & \bf (2000) \rm & \bf (2000) \rm & \bf (sq.deg.) \rm & \bf (Myr) \rm & \bf mass (M$\boldmath _{\odot}$) \rm \\

                 &               &       &        &   & &   \\
(1) IC 4665       & open cl. & 17 46 & +05 43 & 3.1 & 40 & 0.020 \\
(2) Pleiades      & open cl. & 03 47 & +24 07 & 79  & 100 & 0.024 \\
(3) Alpha Per     & open cl. & 03 22 & +48 37 & 50  & 90 & 0.025 \\
(4) Praesepe      & open cl. & 08 40 & +19 40 & 28  & 400 & 0.046 \\
(5) Taurus-Auriga & SF assoc.    & 04 30 & +25 00 & 218 & 1 & 0.010 \\
(6) Orion         & SF assoc.    & 05 29 & $-$02 36 & 154 & 1 & 0.014 \\
(7) Sco           & SF assoc.    & 16 10 & $-$23 00 & 154 & 5 & 0.010 \\
(8) Per-OB2       & SF assoc.    & 03 45 & +32 17 & 12.6 & 1 & 0.011 \\
(9) Hyades        & open cl. & 04 27 & +15 52 & 291  & 600 & 0.041 \\
(10) Coma-Ber      & open cl. & 12 25 & +26 06 & 79  & 500 & 0.043 \\
                &              &       &        &    & &  \\
\end{tabular}

\caption{\label{table:gcs-clus}\it Areas targeted for the 
Galactic Clusters Survey, listed in priority order.
The primary data source is Allen~(1973) with updated 
{\em Hipparcos} distances from Robichon et al (1999) 
and de Zeeuw et al (1999); assuming a depth J=19.5 (Table \ref{table:survey-summary})
the mass limit was calculated using the DUSTY models
of Chabrier et al (2000).}

\end{table*}


\subsection{Design of the Deep Extragalactic Survey (DXS)}

The DXS aims to detect galaxies at
redshifts of 1--1.5.  To avoid
selecting only the brightest and hence most massive galaxies, this
requires the detection of galaxies close to the break in the galaxy
luminosity function, M$^*_{\rm K}$ which is -22.6 locally (Bell \et
2003). At z$=$1 this corresponds to an evolution corrected, total K
magnitude of 20.7, and 21.8 at z$=$1.5. Therefore, taking into account
aperture effects, our target depth of K$=$21 will reach to within 0.5--0.7
magnitudes of M$^*_{\rm K}$ and hence sample a representative galaxy
population at z$=$1. The NIR galaxy colours at this redshift lie in the
range J-K$=$1.5--1.8 so to provide a photometric constraint on the galaxy
redshift we also require observations to J$=$22.3 to ensure matched J and
K detections for the target galaxies.

The survey area was driven by the aim to sample large scale structure at
z$=$1 on scales and volume comparable to that measured locally
($\approx$100~Mpc and 0.2~Gpc$^3$ respectively). At z=1, our assumed
cosmology implies that 100~Mpc corresponds to 3.5$^\circ$ and a
0.2~Gpc$^3$ volume in the range z$=$1--1.5 requires 40 sq.deg. Therefore a
minimum combination of 3x3 WFCAM tiles will span these scales and a total
of 54 WFCAM tiles would be required to cover that area (0.75 sq.deg. per
field).

The number and position of the DXS survey fields were chosen to provide
the best combination of quality and coverage of supporting,
multi-wavelength data, to maximise the spatial scale sampled by each
individual field ($\sim100$~Mpc) for clustering studies and allow a
uniform coverage in right ascension. Balancing these factors resulted in
the selection of four survey fields: 1) XMM-LSS (centre: 2h25m -04d30m) -
a SWIRE, CFHTLS, VVDS, GALEX and XMM survey field adjacent to the UKIDSS
UDS area; 2) the Lockman Hole (centre: 10h57m +57d40m) - centred on the
SHADES survey area but within the SWIRE and GALEX survey areas with
extensive radio coverage; 3) ELIAS-N1 (centre: 16h10m +54d00m)  - a SWIRE
and GALEX field with additional radio, optical and X-ray data; 4) SA22
(centre: 22h17m +00d20m) - centred on VVDS-4 but the least well surveyed
area, included to ensure a uniform demand with RA. Other survey fields were
considered (COSMOS, NOAO-DWFS, Groth Strip, Spitzer-FLS) but most were
either not sufficiently large or comprehensive to justify inclusion. The
chosen fields are listed in Table \ref{dx-uds-fields}.

The total area covered by the DXS within the full 7 year span
of UKIDSS will depend on weather and competition from other
surveys (most notably the UDS) but our goal is 35 sq.deg. or
12 WFCAM fields in each DXS survey area in J and K. 
We also intend towards the end
of the survey to include an additional 1--2 WFCAM fields
in the centre of each DXS survey area in H to broaden the
photometric coverage.

For DXS, star-galaxy separation at faint magnitudes will be very
important, so 2$\times$2 micro-stepping is employed to give good sampling.
Achieving the required depth will require reliable stacking, and so
minimising any systematic effects in detector structure that do not
flat-field out. The DXS strategy therefore employs substantial jittering.
Each visit to a given tile position uses 10 second exposures, a sixteen
point jitter, and 2$\times$2 microstepping at each of these jitter
positions. Each such visit therefore has an integration of 640 second at
each sky point. To reach the intended depth requires a total exposure of
2.1 and 1.5 hours in J and K respectively or 12 and 8 visits each. 
H observations are quite slow, and are planned as a lower priority
in a subset of the chosen fields.
Given
that the DXS observations do not require photometric conditions or the
very best seeing, then the final number of visits for each field may be
higher to compensate for these poorer conditions.

%
\begin{table*}
\begin{tabular}{llllll}
\bf Name \rm & \bf Survey \rm & \bf Area \rm & \bf RA(2000) \rm
 & \bf Dec (2000) \rm   \\

XMM-Subaru      & UDS      & 0.77  & 02 18 00 & -05 10 00   \\
XMM-LSS         & DXS      & 8.75  & 02 25 00 & -04 30 00   \\
Lockman Hole    & DXS      & 8.75  & 10 57 00 & +57 40 00   \\
ELAIS N1        & DXS      & 8.75  & 16 10 00 & +54 00 00   \\
SA22            & DXS      & 8.75  & 22 17 00 & +00 20 00   \\

\end{tabular}

\caption{\it Fields targeted for the Deep Extragalactic Survey and the
Ultra Deep Survey. Note that the UDS field is at the western edge of the
DXS XMM-LSS field}

\label{dx-uds-fields}

\end{table*}


\subsection{Design of the Ultra Deep Survey (UDS)}

The UDS aims to go as  deep as possible in a single contiguous WFCAM tile. The
depth is set by the aim of detecting giant ellipticals at z=3 if they exist. The
total magnitude of such objects is expected to be K$\sim 21$ but they will be
significantly extended, so that we need to reach a point source depth of K=23.
Three bands are needed, to get photometric redshifts and discriminate between objects.
To effectively separate ellipticals and starbursts, we need to be able to detect
colours J-K$\sim 2$ and  H-K$\sim$1, otherwise most of our detections may be
K-band only. This sets limits of J$\sim$25 and H$\sim$24, which are in fact more demanding
in time than the K-band observations. The final expected depths
are J=24.8, H=23.8, K=22.8.

As with the DXS, we need  good sampling to enable star-galaxy separation at
faint magnitudes, and multiple jitters to overcome detector systematics when
stacking. These issues are even more demanding however for the UDS; at each
visit we use 3$\times$3 microstepping and a 9 point jitter, and repeated visits
are not at precisely the same position, but in a carefully arranged pattern - a
kind of super-jitter.

Although the headline science goals of the UDS are to find high redshift 
galaxies which will be extremely faint at optical wavelengths, the legacy 
value of the survey will be increased by the presence of 
complementary ultra-deep optical data. The field chosen was therefore the 
Subaru/XMM-Newton Deep Field (Sekiguchi \et, in preparation) which 
possessed the deepest optical data of any square-degree field at the time 
this decision was made. Considerable data at other wavelengths from radio 
to X-ray also exist in this field. This field is at the western edge of one
of the DXS fields.

\subsection{Two Year Goals}

We aim to complete self contained and scientifically valuable datasets on a 
two year timescale. The detailed plan is set out in Dye \et (2006), but briefly 
is as follows. The shallow surveys (LAS, GCS, and GPS) are accelerated compared 
to the deep stacked surveys (DXS and UDS). In addition, the LAS concentrates
on southern latitudes in the first two years, in order to maximise VLT
follow-up. For LAS, roughly half the area - the equatorial block, and the
southern stripe - will get complete YJHK coverage. (Second epoch J for the 
same areas will come later). Additional J only coverage in the Northern block 
will be achieved as time permits. For GPS, the prime aim is to obtain the first
of three K epochs over the whole survey area, with J, H, and $H_2$ coverage 
over a sub-area. For GCS the aim is to get complete filter coverage for
five of the ten target clusters, and the central regions of three, plus K only
coverage of the remaining two. For DXS, the two year aim is to reach the full
depth in J and K for a subset of the area - four tiles (3.1 square degrees)
in each of the four fields. The UDS is of course a single tile. The two year goal
is to achieve K=22.8 (full depth) and J=23.8 (one magnitude short), 
with no H coverage.

\section{Data Processing and Data Products}  \label{sectn:data}

The commitment to making a  public survey requires the construction of complete,
reliable, tested, and documented products from the raw data. The very large volume of
UKIDSS data (200 GBytes/night) means that to achieve these goals requires a
uniform and automated approach to data processing. Likewise the large
accumulated volume of products - expected to be several tens of Terabytes of image
data and several billion source detections - means that as well as
providing public data access, we need to provide online querying and analysis
facilities as a service.   These ambitions are met for all WFCAM data (both
UKIDSS survey data and PATT PI data) by the VISTA Data Flow System (VDFS). VDFS
is a PPARC-funded project involving QMUL, Cambridge and Edinburgh, aimed at
handling the data from first WFCAM and then the VISTA telescope. (The Science
Archive was also prototyped on the SuperCosmos Science Archive : see
http://surveys.roe.ac.uk/ssa/ and Hambly \et 2004). The system aims at (i) removing instrumental
signature; (ii) extracting source catalogues on a frame by frame basis; (iii)
constructing survey level products - stacked pixel mosaics and merged
catalogues; (iv) providing users with both data access and methods for querying
and analysing WFCAM data. 

Overall data flow is as follows. Raw data are shipped by tape on a weekly basis from Hawaii to Cambridge, where they are available within a month of the observations being taken. Raw data are then transferred via the internet for ingest into the ESO archive system. Pipeline processed single frame data are transferred to Edinburgh over the internet on a daily basis, where they are ingested into the science archive, and further processing (stacking, merging, and quality control) takes place. The processed data are then released to the public at periodic intervals.

\subsection{The WFCAM Pipeline}

The general philosophy behind the pipeline processing is that all 
fundamental data products are FITS multiextension files with headers 
describing the data taking protocols in sufficient detail to trigger the 
appropriate pipeline processing components, and that all derived information, 
quality control measures, 
photometric and astrometric calibration and 
processing details, are also incorporated within the FITS headers.  
Generated object catalogues are stored as multiextension FITS binary tables.  
These FITS files thereby provide the basis for ingest into databases both for 
archiving and for real time monitoring of survey progress and hence 
survey planning. 

After conversion at the summit from Starlink NDF to FITS files, to reduce the data storage, 
I/O overheads and transport requirements, we make use of lossless Rice tile 
compression (eg. Sabbey \et 1998).  For this
type of data (32 bit integer) the Rice compression algorithm typically 
gives an overall factor of 3--4 reduction in file size.  Data are shipped
roughly weekly from JAC using LTO tapes, one per detector channel, and 
combined to create the raw archived multiextension FITS files on ingest in 
Cambridge.

The data processing strategy attempts to minimise the use of on-sky science 
data to form ``calibration'' images for removing the instrumental signature.
By doing this we also minimise the creation of data-related artefacts 
introduced in the image processing phase.
To achieve this we have made extensive use of twilight 
flats, rather than dark-sky flats (which potentially can be corrupted by
thermal glow, fringing, large objects and so on) and by attempting to 
decouple, insofar as is possible, sky estimation/correction from the 
science images.

Each night of data is pipeline processed independently using the master 
calibration twilight flats (updated at least monthly) and a series of
nightly generated dark frames covering the range of exposure times and 
readout modes used during that night.  A running sky ``average'', i.e.
calculated from a number of recent frames, in each 
passband is used for sky artefact correction.  After removing the basic 
instrumental signature the pipeline then uses the header control keywords 
to produce interleaved and/or combined (stacked) image frames for further 
analysis.  This includes generation of detected object catalogues, and 
astrometric and photometric calibration based on 2MASS.  A more detailed 
description of the WFCAM processing is given in Irwin \et (in preparation).

\subsection{The WFCAM Science Archive}

Data processing delivers standard nightly pipeline processed images and
associated single passband catalogues, complete with astrometric and
first--pass photometric calibrations and all associated `meta'
(descriptive) data in flat FITS files. These data are ingested into 
the archive on a more or less daily basis. To 
produce UKIDSS survey products however three more processes 
are needed - image stacking,
source merging, and Quality Control (QC) filtering. 

Source merging is computationally very intensive, UKIDSS catalogues
are extremely large, and the VDFS system is designed to scale to the
even larger expected VISTA products. For standard UKIDSS products
therefore a relatively simple spatial pairing algorithm is used. 
A pairing radius of roughly 1\arcsec (varying between sub-surveys) is used, 
several times larger than the astrometric accuracy (0.1\arcsec), in order to
allow merging of moving objects between epochs. 
Note that the WSA itself follows SDSS practice in storing a {\em neighbour table} 
for every source, so that much more flexible source matching algorithms 
can be applied later, which may be crucial when following up objects with rare 
colours and so on. More details of stacking, mosaicing, and source merging are
given in Irwin \et (in preparation) and Hambly \et (in preparation). 

The QC process is a joint responsibility of the
UKIDSS consortium and the VDFS project and is a semi-automated process. 
The first aim is to remove data that is corrupted or unusable (for
example due to bright moon ghosts). This is achieved partly by automated
inspection of headers and data, and partly by visual inspection. The second
aim is to deprecate a subset of data frames not achieving a standardised `survey quality' 
by applying a series of
cuts based on QC parameters derived from the images - for example, seeing,
average stellar ellipticity, sky brightness and so on. These cuts are evolving
with experience at each data release. Histograms of QC parameters are derived,
and manual decisions taken on the cut levels, which are then applied automatically.
Each sub-survey (LAS, UDS etc) has separate QC parameters. For example, in the EDR,
the LAS seeing limit was 1.2\arcsec, whereas for the UDS it was 0.9\arcsec and 0.8\arcsec at 
J and K respectively. The process is described more fully in 
the ``Early Data Release (EDR)'' paper of Dye \et (2006), and revised parameters
for DR1 are provided in Warren \et (2007a).

Image data volume is typically
$\sim$200~GBytes per night, with catalogue and descriptive data being
typically $\sim10$\% of that figure. Hence, over the course of several
years of observations it is anticipated that 10s/100s of Tbytes of
catalogue/image data will be produced by survey operations with WFCAM.
In order to enable science exploitation of these datasets, the concept
of a `science archive' has been developed as the final stage in the
systems--engineered data flow system from instrument to end--user
(Hambly \et~2004b).

Finally, we note that the WSA has been developed throughout
with Virtual Observatory compatibility in mind. 
The WSA exports VOTABLE, exposes its query interface
as an ADQL web service, is published in VO registries, and provides access
to AstroGrid VO tools such TopCat and MySpace.

%
\begin{figure}

\includegraphics[width=85mm,angle=0]{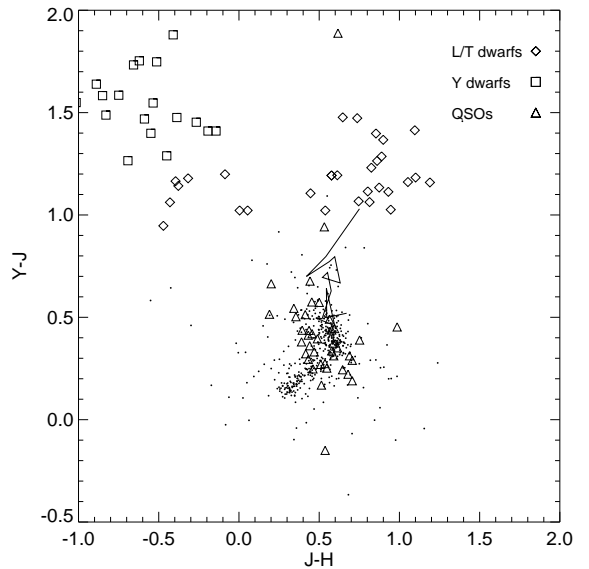}

\caption{\label{fig:udef-sv-las1} \it
The Y-J,J-H two colour diagram for a single tile observed
in the LAS SV programme. Black dots show the data for 
stellar sources detected in the WFCAM data. 
Also shown are the synthetic 
colours of QSOs, L/T dwarfs, and model Y dwarfs.
The solid line shows the positions of M dwarfs. The observation was 
targeted at a known T2 dwarf,SDSS J125454-012247 which is recovered with
Y-J=1.10 and J-H=0.54.}

\end{figure}
%
%
%
\begin{figure}

\includegraphics[width=150mm,angle=-90]{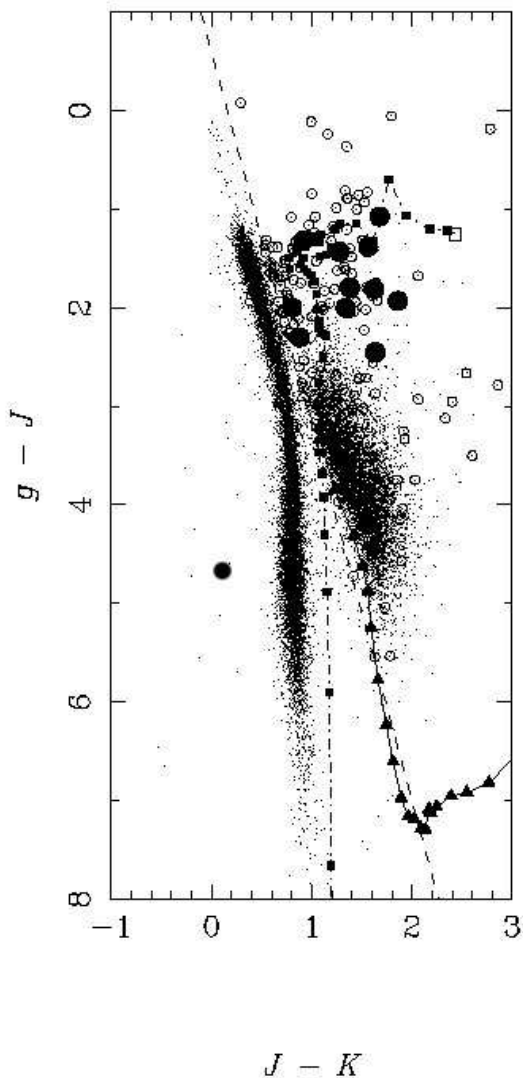}

\caption{\label{fig:udef-sv-las2} \it
Illustration of the KX method from some 10 sq degs of science  
verification observations in high Galactic latitude fields. All  
detected objects in the range 14.5$<$K$<$16.5 are plotted, totaling  
21000 sources. Stars make up the long, thin cloud, and galaxies form  
the shorter cloud to the right. The solid squares are model quasar  
colours 0$<$z$<$8, $\Delta$z=0.1, from Hewett et al. (2006), with z=0  
marked by the open square. Similarly triangles mark the model colours  
of an unevolving elliptical galaxy 0$<$z$<$3, $\Delta$z=0.1. The  
large filled circles are the 11 SDSS spectroscopically confirmed  
quasars in the observed fields, brighter than the fainter limit of  
K=17. The diagonal dashed line represents a possible KX selection  
criterion. Candidate quasars (of which there are 167) are compact  
sources to the right of the line, and are indicated by the open  
circles. Reddening vectors at different redshifts run approximately  
parallel to this line (Warren et al. 2000).
}

\end{figure}

%
The WFCAM Science Archive\footnote{http://surveys.roe.ac.uk/wsa} (WSA) 
is much more than a simple repository
of the basic data products described previously. A commercial
relational database management system (RDBMS) deployed on a mid--range,
scalable hardware platform is used as the online storage into which
all catalogue and meta data are ingested. This RDBMS acts as the backing
store for a set of curation applications that produce enhanced database
driven data products (both image products, e.g.\ broad--band/narrow--band
difference images; and catalogue products, e.g.\ merged multi--colour,
multi--epoch source lists). Moreover, the same relational data model is
exposed to the users through a set of web--interface applications that
provide extremely flexible user access to the enhanced database driven
data products via a Structured Query Language interface.
The primary purpose of the WSA is to provide user access to UKIDSS
datasets, and a full description, along with typical usage examples,
is given in Hambly \et~(in preparation). Step-by-step
examples of WSA usage are also included in the UKIDSS EDR paper, Dye \et (2006).

\section{Science verification programme}  \label{sectn:sciver}

%
\begin{figure}

\includegraphics[width=85mm,angle=270]{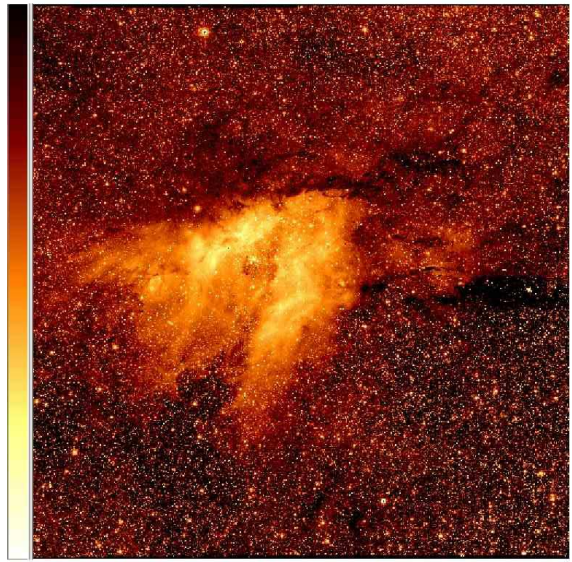}

\caption{\label{fig:udef-sv-gps1} \it
Central 25\% of a GPS K-band tile pointed at M17, showing the richness of data in the 
GPS. The full tile contains 740305 sources.}

\end{figure}

%

Because the consortium has no proprietary rights, and 
no formal role in deriving science results from the data
(as opposed to facilitating their exploitation), in this
paper we have avoided showing science results from the
released data. Examples of early papers using UKIDSS
data are given in section \ref{sectn:first-util}.
However, the potential of UKIDSS can be
well illustrated using the Science Verification data.

Following the technical  commissioning of WFCAM, and before the commencement of
formal survey operations in May 2005, the UKIDSS consortium undertook a modest
set of test observations, as a ``Science Verification (SV)'' programme.
These observations were aimed primarily at further technical
commissioning, testing and tuning the implementation strategy, and exercising the data flow
system. However the data collected have clearly demonstrated the scientific
power of UKIDSS, and confirm the efficacy of the survey design. In this section
we show some examples of science results from these SV data.

\subsection{Science Verification results for the LAS}

The LAS SV programme covered some 20 square degrees, achieving
close (0.2mag) to the 
standard shallow depth in filters Y,J,H,K. 
One scientific aim was to test the likely recovery of cool brown
dwarfs. The success of this is illustrated in Fig. \ref{fig:udef-sv-las1}, which 
shows data from a tile aimed at a known T2 dwarf, 
SDSS J125454-012247 (Knapp \et 2004).
This object was indeed detected, and the colours found are consistent with 
those previously published. Other L and T dwarfs were also targeted and 
successfully detected. In this very limited area no new objects of 
significant interest have been found but that is in line with 
expectations. The LAS SV data were also used to verify the photometry and 
astrometry of the UKIDSS survey. Details of these tests
are given in the EDR data release paper, Dye et. al. (2006).

A second aim was to test the location of quasars by combining
UKIDSS and SDSS colours. Figure \ref{fig:udef-sv-las2} 
shows that the ``K excess'' method works extremely well. The stellar
and galaxy sequences are cleanly distinguished. 
Point-like objects to
the right of the dashed line are good quasar candidates. 
Known SDSS quasars
in these fields are in the upper part of this region, but the UKIDSS
SV data shows many more candidate quasars with similar colours.
Additionally there are several candidates with much  
redder colours than the known quasars. These are candidate reddened  
quasars, and spectroscopy is required to investigate their nature.  
Several have colours similar to galaxies, but the overall colour  
spread of the candidates is much broader than for the galaxies. 
The quasars in these fields, if confirmed, will be at relatively modest
redshift. Note  
that quasars move rapidly redder in $g$-J beyond z$\sim$3.8.
The very high redshift quasars that we hope to find will be much
sparser on the sky, and in gJK will be hard to distinguish from cool brown dwarfs.
Here, as explained in section \ref{sectn:goals-las}, the Y-J colour will
be crucial. Analysis of relevant data is still in progress
and will be reported in a later paper.

\subsection{Science Verification results for the GPS}

%
\begin{figure*}

\includegraphics[width=70mm,angle=0]{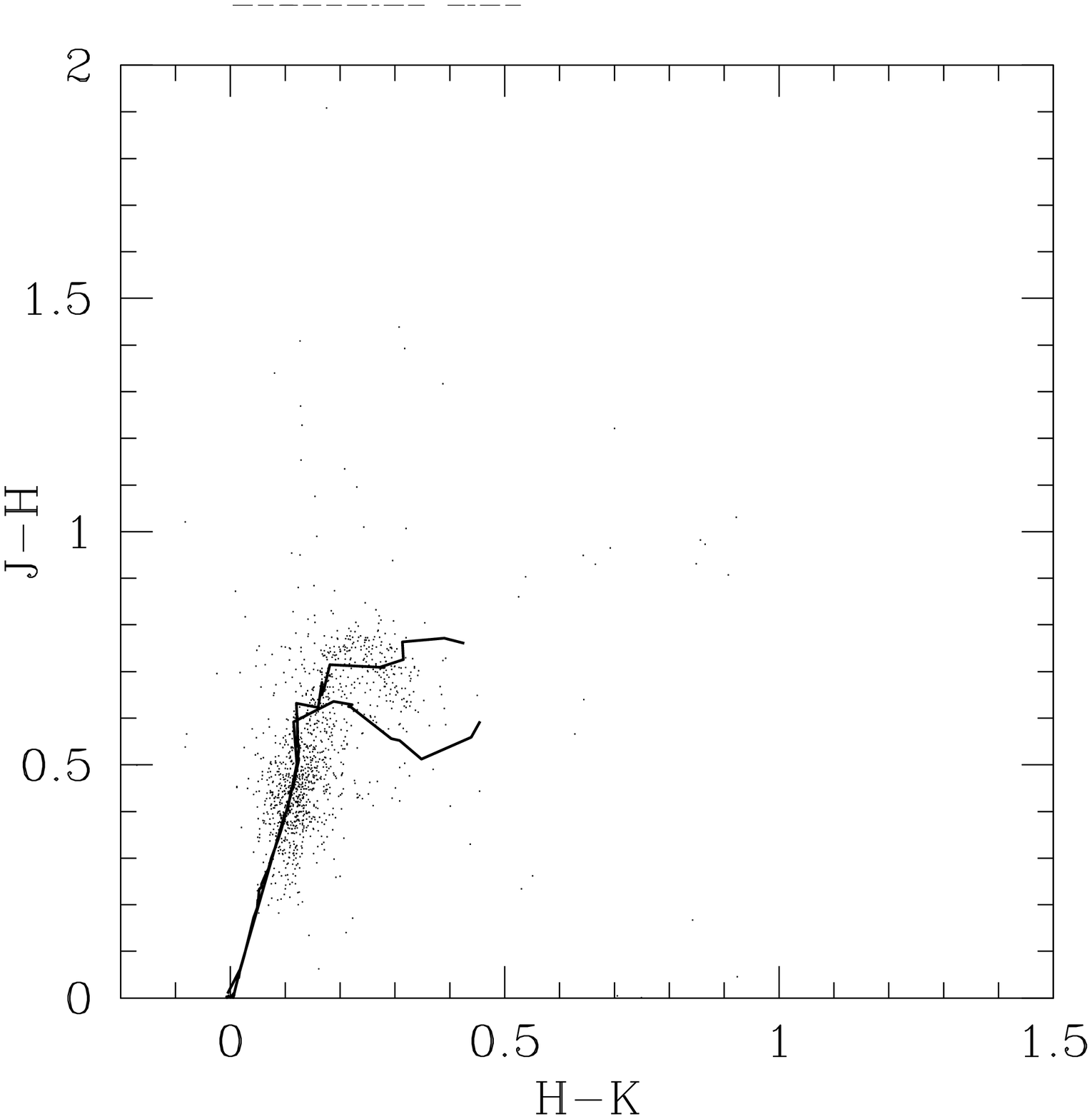}
\includegraphics[width=70mm,angle=0]{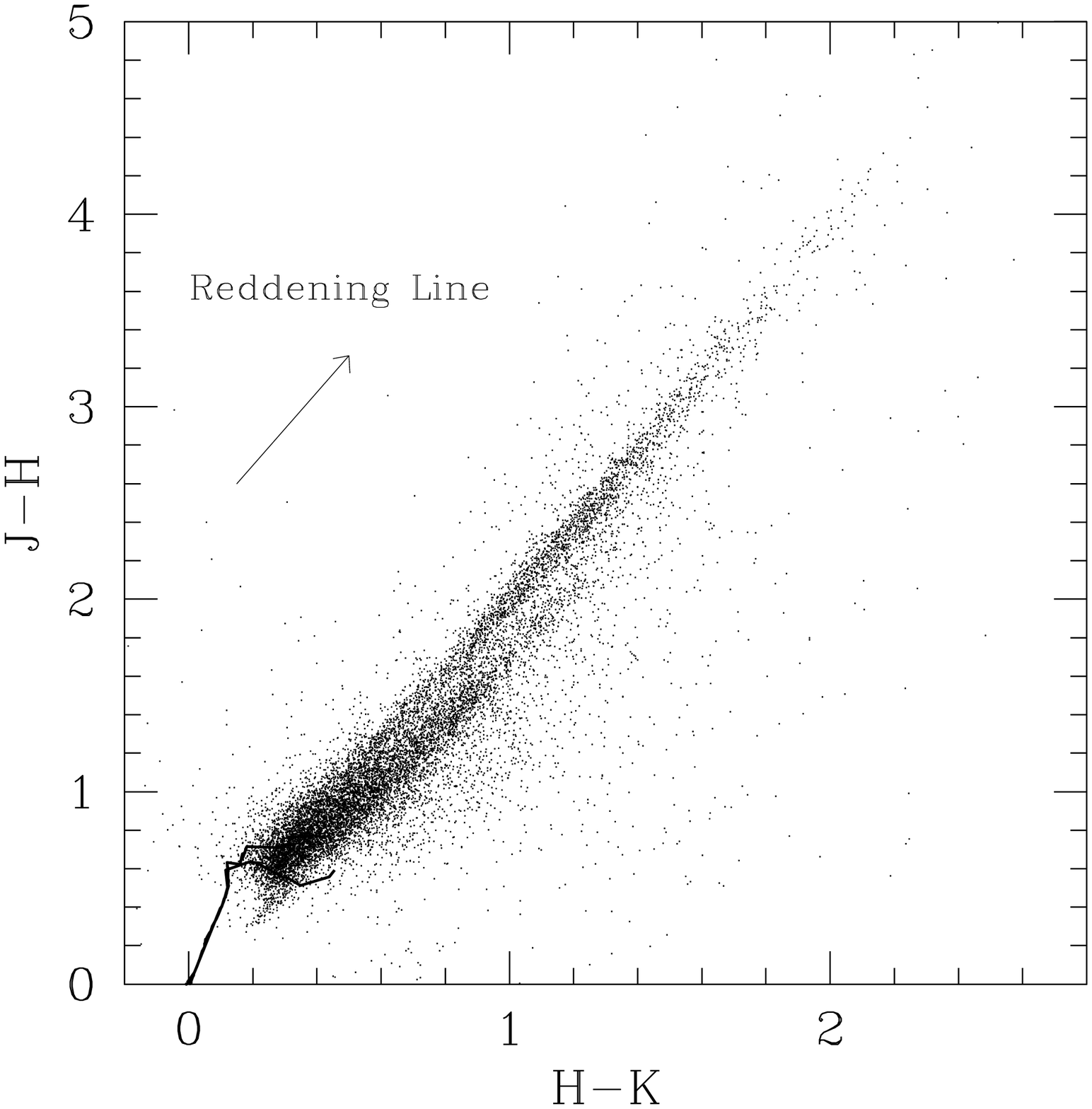}

\caption{\label{fig:udef-sv-gps2}
\it Two colour diagrams for two individual array detectors
(12.8 arcmin fields), showing sources listed as having
errors $<$0.05 mag on each axis in the WSA. {\bf Left hand panel} :
an uncrowded field with low extinction at ($l,b$)=171.2,4.8,
showing 1236 sources.
{\bf Right Hand panel} : a crowded field with high extinction at 
($l,b$)=15.0,0.0, showing 12954 sources. In each panel the lower 
curves shows the locus of 
the main sequence, and the upper curve shows the locus of luminosity 
class III giants. Even in the crowded field the photometric precision
is good enough to show the 0.25 mag vertical separation between the 
reddened giant sequence and the reddened main sequence stars. An 
approximate reddening vector is plotted in the right hand panel. Some
curvature of the reddening sequences is seen in the data, which is 
attributed to the change in the effective bandpasses at high
extinction.
}

\end{figure*}

%

The science verification data for the Galactic Plane Survey included a
0.75~deg$^2$ tile centred on M17 (a high mass star formation region). 
The M17 region has been well studied (eg. Jiang \et 2002) and it provided 
a good test of the photometric reliability of the data in a nebulous region.

The central 25\% of the M17 tile is shown in Figure \ref{fig:udef-sv-gps1} 
This illustrates the sensitivity of WFCAM to the structure and stellar
population of distant star formation regions. By visual comparison
of the WSA catalogue and the reduced image, it is clear that
very few spurious sources are 
found in the archive but within the brightest
nebulosity the archival source lists become seriously incomplete, with 
almost all point sources being undetected.
This problem occurs generally throughout the GPS in very bright nebulae,
since the UKIDSS pipeline photometry is designed to detect resolved
galaxies as well as point sources, and hence it attempts to interpret 
bright nebulosity as a single extended source. Such very bright
nebulae represent a tiny fraction of the total survey area.
In general, more complete luminosity functions may be derived in nebulous 
regions by performing independent photometry on the reduced data, using the
zero points provided in the image headers.

At the present time only aperture photometry is available in the
WFCAM Science Archive. One might expect the photometric precision to 
suffer considerably from the effects of source confusion. However the algorithm 
simultaneously fits the coordinates and fluxes of adjacent stars in small groups
and divides up the fluxes between any overlapping apertures in a sensible
manner (following principles described by Irwin 1985). Hence the resulting
photometric quality is very good even in very crowded fields with source
densities in excess of 10$^6$ per square degree, and is similar to the quality
of results from profile fitting photometry, which is a far more time consuming 
process. (Detailed comparison of various photometry techniques will be given in Irwin \et\ (in preparation)).

The quality of the photometry is illustrated in the two colour-colour diagrams
in Figure \ref{fig:udef-sv-gps2}, each of which shows those sources within
a single detector array (13 arcminute field) that are listed as having well measured 
photometry (errors $<0.05$~mag on each axis) in the WSA. 
The curves shown are for unreddened luminosity class V main sequence stars 
(lower curve) and 
luminosity class III giants (upper curve), using synthetic photometry from Hewett 
\et (2006). The left hand panel shows a field at $(l,b)$=171.2, 4.8, where there is 
very little extinction. Nearly all of the stars lie near the curves, and both
the G-type to early M type main sequence and the giant branch are well
represented. There are relatively few sources with colours 
consistent with those of late M dwarfs near the end of the class V sequence. Hence 
bona fide late M and L dwarfs should be detectable by follow up observations with a
reasonable success rate, provided they have precise photometry. The task will
become easier when proper motion information becomes available with the 2nd and 
3rd epoch data.

The right hand panel of  \ref{fig:udef-sv-gps2} illustrates a crowded field with high 
extinction at $(l,b)$=15.0, 0.0. Despite significant source confusion, there
is a clear separation between the reddened giant branch and the reddened
main sequence. The majority of the stars with low reddening are K-type to 
early M type dwarf stars. Giant stars are generally more distant and suffer
higher reddening, so the giant branch becomes clearly defined only at 
$(J-H)>1.2$, $(H-K)>0.6$. Some curvature of the reddening sequences
is apparent, which we attribute to the change in the effective wavelengths
of the JHK bandpasses at high reddening. 
The number of photometric outliers can be greatly reduced by using 
additional quality information in the archive, eg. using source 
ellipticity to remove binaries which are unresolved in one or two of the 
passbands. This will be explored fully in a later paper.

Since the main sequence curve is almost parallel to the reddening vector it 
appears that additional colors will be required to permit precise photometric
determinations of source extinction and hence spectral type and 
luminosity class. For blue sources this will be done with the aid
of optical data from the IPHAS survey (www.iphas.org) while 
for very red sources mid infrared 
data from the SPITZER-GLIMPSE survey (www.astro.wisc.edu/sirtf/) of part 
of the galactic plane and later the NASA WISE survey of the whole sky 
(wise.ssl.berkeley.edu/news.html.) may be useful.

Profile fitting photometry will be included in a future release of UKIDSS data 
if it is found to improve the photometric precision. Experiments with
DAOPHOT in IRAF and a preliminary profile fitting scheme for the microstepped 
data (D. Wyn Evans, private comm.) indicate that it is unlikely to produce a 
major improvement over aperture photometry but it may be possible to 
reduce the number of photometric outliers and thereby aid the detection
of sources with unusual colours.

In relatively uncrowded regions the aperture photometry reaches close to the 
target depths at J, H and K (see results in Warren \et 2007). Source confusion
does not significantly affect the measured depth (K=18.1) in the outer galaxy section
of the GPS at $l$=142-230. Even in the inner galaxy section ($-2<l<107$) the 
reduction in sensitivity is $<$0.5 mag in most fields. However the sensitivity suffers
more in the most crowded parts of the mid-plane as the Galactic Centre is approached, 
eg. it is reduced by 1.5 mag at $l$=15, $b$=0.

A full analysis of the data quality in the GPS and some results from
the SV data will be presented in a future paper (Lucas \et in preparation).

%
\begin{figure}

\includegraphics[width=75mm,angle=0]{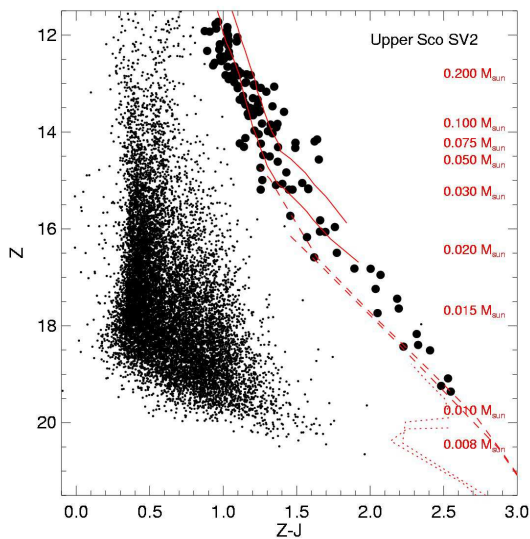}

\caption{\label{fig:udef-sv-gcs1} \it
Z vs Z--J diagram for six square degrees in the Upper Sco field, 
showing only those sources
with  J$>$10.5 and Z$>$11.5. 
Candidate cluster members have been isolated by colour cuts in 
multicolour space (see text).
One in ten of the field sources are plotted, and all the cluster member
candidates.
Model isochrones are described in the
text.
The cluster sequence can be clearly seen following the models
well. Objects on or to the right of the model isochrones, and so likely to
be cluster members, are plotted
as large points. The expected positions of objects of various masses
are indicated.  Sub-stellar objects are clearly detected.}

\end{figure}
%

\subsection{Science Verification results for the GCS}

Science verification observations for the GCS yielded 8 tiles in each of the
three targets observable at that time: IC~4665, Upper Scorpius and Coma
Berenices. In the case of IC~4665, the SV observations complete the required
survey for that cluster. In Figure~\ref{fig:udef-sv-gcs1} we show a Z versus 
Z--J  colour--magnitude diagram for Upper Scorpius. The observations cover
6 square degrees and approximately 100,000 point sources are detected. 
Nearly all of these are of course background stars, with perhaps some foreground stars. 
In Fig \ref{fig:udef-sv-gcs1} we
show only those sources with  J$>$10.5 and Z$>$11.5. The main sequence 
and giant branch are clearly seen. In addition one can notice a clean
sequence to the right of the diagram running from (Z-J,Z)=(1.0,12.0)
to (Z-J,Z)=(2.5,19.0), which must be the cluster sequence. Candidate cluster
members, plotted as large points in Fig. \ref{fig:udef-sv-gcs1}, 
have been isolated first by selection in the Z-J vs Z diagram,
and then in Y-J vs J-K to eliminate field dwarfs. Objects brighter than
J=15.8 are also detected in 2MASS, giving a proper motion estimate
which allows a very reliable final determination of cluster membership.
For fainter objects we have only the photometric information, but
estimate that field contamination is less than 10\%. (See 
Lodieu \et (2007) for details). 

Overplotted are 5 Myr theoretical 
isochrones shifted to a distance of 145pc, appropriate for 
the estimated age and distance of Upper Sco : BCAH98 or NextGen models (solid line; Baraffe et 
al. 1998), DUSTY or BCAH00 (dashed line; Chabrier \et 2000), and COND03 
(dotted line; Baraffe \et 2003). These isochrones were 
specifically computed for the WFCAM filters 
(Isabelle Baraffe and France Allard; personal communication).
On these isochrones we also
indicate object masses. We can clearly see therefore that we are indeed
locating brown dwarfs within the cluster and will be able to derive a
substellar mass function all the way down to 10 Jupiter masses.

\subsection{Science Verification results for the DXS and UDS}

\begin{figure}

\includegraphics[width=80mm,angle=0]{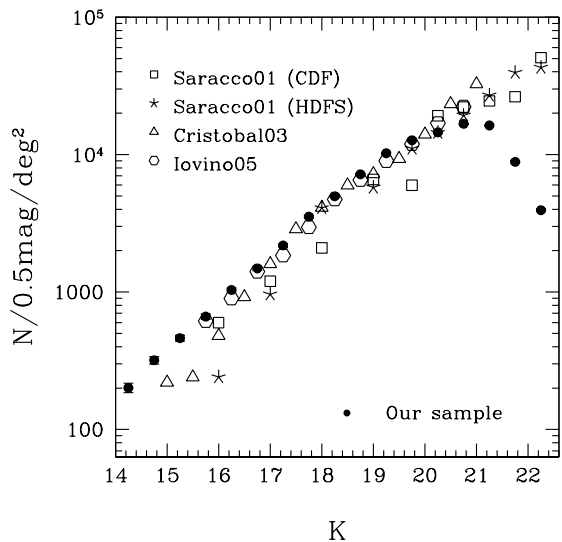}

\caption{\label{fig:udef-sv-uds1} \it
K-band number counts from six hours of UKIDSS DXS/UDS 
SV observations of a single tile
in the ELAIS N1 field. Results from the literature are
shown for comparison. Note that the UKIDSS number counts include
both stars and galaxies. }

\end{figure}

The DXS and UDS undertook a joint science verification
programme to establish the performance of WFCAM for deeper
exposures.
Fig \ref{fig:udef-sv-uds1} shows the K-band number counts from an accumulated
exposure of 1.5 hours (6 hours of data) on a single tile in the ELAIS-N1 field.
Simulations show that the 50\% completeness limit in this field is K=20.3, significantly shallower than the expected 5$\sigma$ limit in this time.
This is partly but not wholly because of poorer than 
average seeing in the SV observations. The reason for this discrepancy is
discussed in section \ref{sectn:survey-progress}.
Note that this figure includes both stars and galaxies. For comparison, we show
number counts from several other surveys. This figure illustrates the power
of UKIDSS, as we can determine number counts accurately from a single
tile. When the survey is complete, we will therefore be relatively immune to cosmic variance.

\begin{figure}

\includegraphics[width=80mm,angle=0]{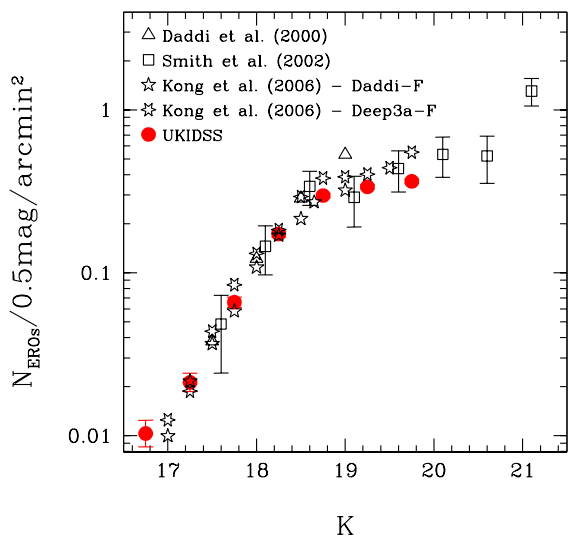}

\caption{\label{fig:udef-sv-uds2} \it
Number counts of Extremely Red Objects (EROs), taken
from the same 
DXS/UDS SV field as in Figure \ref{fig:udef-sv-uds1}, but
selected to have  $R-K>5.0$. These are
compared with ERO counts from the literature, although note that Smith et
al. (2002) use a slightly different selection criterion ($R-K>5.3$).
}

\end{figure}

One of the scientific goals of DXS and UDS is to locate Extremely Red Objects (EROs).
Again, with a tiny fraction of the eventual survey data we are already
competitive with all previous studies. Using the same UKIDSS SV data as above,
we cross-match with publicly available optical data from the INT Wide Field
Camera Survey\footnote{http://www.ast.cam.ac.uk/$\sim$wfcsur/}, 
and select EROs as those with R-K$>$5 down to a limit
of K=19. This produces a sample of 1660 EROs. The reliability of this sample
is limited by the optical data at R=24, not by the IR data at K=19.
Fig. \ref{fig:udef-sv-uds2} shows the number counts of EROs in these data. Again,
we have already duplicated most previous work in these science verification data.

\section{Survey Progress}  \label{sectn:survey-progress}

The final version of this paper is being written in April 2007. We have made
an `Early Data Release (EDR)' (Dye \et 2006), and First and Second Data Releases 
(DR1 and DR2;  Warren \et 2007a,b).
The EDR and DR1 papers contain considerable detail on the areas covered,
data quality achieved, QC cuts applied, example SQL queries for extracting 
science ready data, and so on, with the DR2 paper providing an incremental update.
DR1 contains approximately 7\% of the expected final dataset,
with some surveys proceeding faster than others. Table \ref{table:survey-progress} summarises
the content of DR1.

%

%

%

%

\subsection{Survey quality achieved}  

After filtering data through the QC process, UKIDSS is achieving its design goals,
with the partial exception of depth in stacked surveys, which is expected to improve
in later releases. Except where noted, the statistics quoted below refer to measurements made from the DR1 database. 
Distributions of these quantities are shown in Warren \et (2007a).

{\bf Image Quality.} The median seeing in DR1 is $0.82^{+0.15}_{-0.13}$\arcsec , and 
the median stellar ellipticity is $0.07^{+0.04}_{-0.02}$ (1$\sigma$ ranges). These figures
are expected to improve slightly in later releases ; nights used in the first half 
of DR1 had imperfect alignment of the optics and flexure correction. In QC filtering, 
seeing cuts were applied to the various surveys, roughly as indicated in 
Table \ref{table:survey-goals}.

{\bf Photometric Accuracy.} Calibration uses 2MASS stars over the whole of each frame, and 
the internal dispersion indicates that locally we are reaching the target 
accuracy of 0.02 mag RMS., and that frame gradients are negligible.  We also have compared the 
average colours of stars from field to field, which gives colour accuracies of 0.02 -- 0.03 mag, 
indicating that photometric errors are less than 0.02 mag in all filters.
The global uniformity is currently as good as the 2MASS survey itself 
allows, which is believed to be 2\% (Nikolaev \et 2000). As well as calibrating from 2MASS, we 
are collecting observations of UKIRT faint standards, with which we will check uniformity at faint magnitudes (Hodgkin \et in preparation). All the above applies to an average star in an average field.
The analysis of DR1 found some small but clear variation depending on stellar type and local extinction, indicating that some improvement is needed in colour transformations between the 2MASS and UKIDSS
systems. These will be applied in DR2.

{\bf Astrometric Accuracy.} Astrometric accuracy was considered in some detail in the EDR (Dye \et 2006).  As with photometry, the astrometric calibration is tied to 2MASS stars, which in turn are tied to the Tycho system. The global uniformity therefore assumes that of 2MASS, although this will in due course be checked independently. The precision of the astrometric solution changes with Galactic latitude, because of
the number of 2MASS stars available for solution. The RMS accuracy varies from 50mas at low latitudes to 100mas at high latitudes.  

{\bf Depth.} The depth achieved in the shallow surveys is close to expectation. From LAS data in DR1, the median 5$\sigma$ point source depth in the standard 40 second shots, compared to target depths is Y=20.16 (20.3), J=19.56 (19.5), H=18.81 (18.6), and K=18.19 (18.2). From the GCS, the Z depth achieved is Z=20.36 (20.4). For most surveys, this is a good guide to actual achieved catalogue depth, but in due course this will be analysed more fully in terms of completeness and relaibility. For the GPS in particular, the effective depth is 
very sensitive to confusion, and so varies with position on the sky (see Warren \et 2007). 
For the stacked surveys, experience so far has borne out what we found in the SV data - the depth achieved falls somewhat short of the naive $\sqrt N$ extrapolation. In the current pipeline output for the DXS, we are a few tenths of a magnitude short of target. 
At the eventual depth of the UDS, K=23, we expect of the order 100 galaxies per sq. arcmin, 
which should produce significant correlated confusion noise, as seen in other very deep IR 
surveys, such as the ISAAC FIRES survey of Labb\'e \et (2003). However, we are still some way short 
of this depth. The reason for the stacked survey depth discrepancy is currently unclear.
It is possible that pipeline improvements and more careful analysis will remove much 
of this discrepancy. Further analysis will be provided in later data release papers.

{\bf Database utility.} The VDFS pipeline and archive system has delivered more or less what was required for UKIDSS, including reduced data quality and the ability to perform catalogue queries online.  There are known processing limitations, listed on the WSA web page. These are relatively minor, with two main exceptions. The first is that profile magnitudes have not yet been implemented, which is potentially important for GPS analysis. The second is that deep stacking is not yet optimal. For this reason, the UDS working group have created their own image stack and deep catalogue outside the VDFS system; however this has been supplied to the WSA and is currently used as the publicly available product. Both of these issues will be resolved for later releases. 

{\bf Reliability and completeness.} We have so far made only a preliminary study of completeness as a function of depth, or the number of spurious sources, for the various UKIDSS products. This is partly because they are {\em databases} rather than statistically defined catalogues. A number of different catalogues can be extracted from the databases, defined with different cuts on quality parameters etc. No single catalogue is perfect - rather these should be extracted by users of the UKIDSS databases for specific scientific purposes. However, as the UKIDSS consortium, we will provide the best possible quality flags, and in due course we will produce an analysis of representative catalogues as a guidance to users. Given that many scientific uses of UKIDSS will involve looking for rare or anomalous objects, reducing contamination by spurious sources due to data artefacts, and wrong colours due to incorrect band-merging, are both crucial. The frequency and nature of artefacts is discussed in Dye \et (2006) and Warren \et (2007), and source merging is discussed in detail in Hambly \et (in preparation), but we have not yet estimated the density of spurious sources or incorrect mergers. Source variability can also mimic unusual colours, given that some passbands are measured at separate epochs. This can only be assessed by attempting the relevant science projects. When this issue clarifies, later data release papers will include an estimate of the problem.

%
\begin{table}
\begin{tabular}{lcccc}
\bf Survey \rm & \bf Area deg$^2$ \rm  & \bf Filters \rm  
& \bf K-depth \rm  & \bf Compn \rm \\

               & \bf all (any) \rm    &     & \bf 5$\sigma$ Vega \rm & \\
               & \bf filters \rm    &     &  & \\
               &   &     &  & \\

LAS  & 190 (475)  & YJHK  & 18.2 & 6\% \\
GCS  &  52 (401)  & ZYJHK  & 18.2 & 13\% \\
GPS  & 77  (362)  & JHKH$_2$  & 18.1 & 7\% \\
DXS  & 3.1  (6.9)  & JK  & 20.7 & 11\% \\
UDS  & 0.8 (0.8)  & JK  & 21.6 & 4\%  \\

\end{tabular}

\caption{\it Summary contents of the UKIDSS First Data Release (Warren {\et} 2007).
Depth achieved is shown only for the K-band, and the `all' area is that
with data achieving that depth and covered in all relevant bands. 
The completion estimate takes account of all filters, and areas
with partial depth.}

\label{table:survey-progress}

\end{table}


\subsection{Survey completion rate}  

Up to DR1 release, the rate at which the survey has accumulated has been slower than hoped. Raw data has been collected on average at 79\% of the rate planned, and in addition 20\% of data frames have failed our QC filtering, so that the net survey progress rate has been 63\% of our goal. 
Since completion of DR1 observing, two major problems - bright moon ghosts, and an unexpectedly high data acquisition system overhead - have been solved. Our current estimate is that the net survey progress rate is now $\sim$80\% of goal. Within the first two years of UKIDSS, we expect to have completed approximately 2/3 of our two year plan. To achieve the full design parameters of Table \ref{table:survey-summary} will take approximately 1150 nights. The fraction of UKIRT time dedicated to WFCAM is under review, so the number of years required, and/or the adjusted survey parameters, are currently uncertain. 

\subsection{First utilisation of UKIDSS}  
\label{sectn:first-util}

In section \ref{sectn:sciver} 
we have shown analyses undertaken by the consortium with science verification data. Since EDR and DR1, the first open community exploitation of UKIDSS has begun. Use of the UKIDSS databases in the WSA has been extensive. As of the end of 2006, over a billion rows of catalogue data have been downloaded, and interesting results are starting to emerge. 
The first very high redshift quasar (z=5.86) has been found by Venemans \et (2007). McLure \et (2006) have reported the discovery of nine of the most luminous known Lyman break galaxies, at $5<z<6$ - they are surprisingly massive so soon (1.2 Gyr) after the Big Bang. A sample of 239 Distant red Galaxies (DRGs) at $2<z<3$ has been selected by Foucaud \et (2007) and their angular correlation function measured. Cirasuolo \et (2007) have measured the evolution of the K-band galaxy luminosity function to $z\simeq 2$. The coolest known brown dwarf (classified T8.5) has been found by Warren \et (2007c; spectrum shown in Warren \et 2006). Lodieu \et (2007) have found 129 new brown dwarfs, a significant fraction of the total known, including a dozen below 20 Jupiter masses. Large new samples of quasars, distant galaxy clusters, and extremely red objects have been constructed by Chiu \et (2007), Van Breukelen \et (2006), and Simpson \et (2006) respectively.

\subsection{Survey releases}  \label{releases}

Data access policy for UKIDSS is set by the UKIRT Board, and is set out
on the JAC web pages\footnote{http://www.jach.hawaii.edu/UKIRT/surveys\newline /UKIDSSdatapolicies.html}. UKIDSS is intended to produce multi-use data of general benefit 
to astronomers worldwide, but with a temporary advantage for the communities 
that developed the camera and surveys. Initially this meant UK astronomers, but 
now means any astronomer currently working in an ESO member state. 
The general principle is that the data are freely available to any such astronomer
from the point of release, and available worldwide eighteen months later. (Note
that individual members of the consortium have no privileged data access.)
During the ESO-restricted phase, data access requires registration with the
WSA. This is organised through a set of ``community contacts'' at astronomical
institutions in ESO member states, who maintain their own databases of local users
through the WSA system. Any reader who is not yet registered who believes they are eligible
should contact their local community contact, or if necessary ask for a new community contact
to be established. Fuller instructions and a list of current contacts is on the UKIDSS web page (http://www.ukidss.org/archive/archive.html).

It is intended to make UKIDSS data available in a series of well defined staged releases.
As well as involving incrementally more data, releases will usually 
involve reprocessing of data from previous releases, 
with updated correction of artefacts and so on. 
Each release will therefore be documented by a paper describing the contents 
and limitations of that release. The first preliminary release, avalable from Feb 10th 2006,
had a relatively small amount of data (about 1\% of the expected total), 
and several known imperfections in the data processing. 
This was therefore labelled an ``Early Data Release (EDR)''. It is described in more detail
in Dye \et (2006). The first full data release (DR1) took place on July 21st 2006. It contains
7\% of the intended final data volume, and has substantially better data quality than the EDR in a variety of ways. It is described in Warren \et (2007a). At the time of final revisions to this paper,
the Second Data Release (DR2) has also recently taken place, with updated details in an online-only paper (Warren \et 2007b). 

Further releases are likely to take place thereafter every six months or so. 
Raw WFCAM data 
are available through the ESO archive system (http://www.eso.org) and
through the CASU site (http://archive.ast.cam.ac.uk/wfcam/). 
All of the UKIDSS processed images and catalogues are accessible and queryable through
the web-based WFCAM Science Archive (WSA : http://surveys.roe.ac.uk/wsa).


\section{Acknowledgements}  \label{acknow}

This paper is written on behalf of the entire UKIDSS consortium. 
The membership list can be found at
http://www.ukidss.org. The formal
authorship includes the heads of the various UKIDSS working groups, plus
a small number of other individuals. 
In addition to the bulk of the consortium,
there are others we would like to thank.
First and foremost, the UKIDSS enterprise would be impossible without
the staff of UKIRT, the staff at UKATC who built WFCAM, and the staff
of CASU and WFAU who built the data processing system. Next, we would like to
note that much of UKIDSS, including its scientific ambition, design, data flow concepts, and release plan, has been built upon the preceding ideas and high professional standards of the SDSS and 2MASS teams. Finally we would like to thank the anonymous referee, whose detailed and thoughtful comments improved this paper significantly.


\section{REFERENCES} \label{refs}

\noindent Allen, C.W., 1973, Astrophysical Quantities, 3rd edition 
(University of London; Athlone)
\smallskip

\noindent Baraffe, I., Chabrier, G., Allard, F., Hauschildt, P. H. 1998, A\&A, 337, 
403
\smallskip

\noindent Baraffe, I., Chabrier, G., Barman, T.S., Allard, F., Hauschildt, P. H. 2003, A\&A, 402, 
701
\smallskip

\noindent Bell, E.F., McIntosh, D.H., Katz, N., Weinberg, M.D. 2003 ApJS 149, 289
\smallskip

\noindent
van Breukelen, C., Clewley, L., Bonfield, D.G., Rawlings, S., Jarvis, M.J., Barr, J.M., 
Foucaud, S., Almaini, O., Cirasuolo, M., Dalton, G., Dunlop, J.S., Edge, A.C., 
Hirst, P., McLure, R.J., Page, M.J., Sekiguchi, K., Simpson, C., Smail, I., 
Watson, M.G., 2006, MNRAS, 373, L26.
\smallskip

\noindent Burrows, A., Sudarsky, D., and Lunine, J.I., 2003, ApJ, 596, 587.

\noindent Casali, M.M., Adamson, A.A., Alves de Oliveira, C., \et 2007, A\&A 467, 777
\smallskip

\noindent Chabrier, G., Baraffe, I., Allard, F.,  Hauschildt, P. H. 2000, ApJ, 542, 
464
\smallskip

\noindent Chiu, K., Richards, G.T., Hewett, P.C., Maddox, N., 2007, 
MNRAS, 375, 1180 (astro-ph/0612608)
\smallskip

\noindent Cristobal, D., Prieto, M., Balcell, M., Guzman, R., Cardiel, N.,
Serrano, A., Gallego, J., Pello, R. 2003, Science with the GTC
Eds. J. Espinosa, F. Lopez, and V. Martin) Revista Mexicana de
Astronomia y  Astrofisica (Serie de Conferencias) Vol. 16, pp. 267
\smallskip

\noindent Cirasuolo, M., McLure, R.J., Dunlop, J.S., Almaini, O., Foucaud, S., 
Smail, I., Sekiguchi, K., Simpson, C., Eales, S., Dye, S., Watson, M.G., 
Page, M.J., Hirst, P., 2007,  MNRAS submitted (astro-ph/0609287)
\smallskip

\noindent Daddi, E., Cimatti, A., Pozzetti, L., \et 2000, A\&A, 361, 535
\smallskip

\noindent Dalton, G.B., Calswell, M., Ward, K., \et 2004, SPIE, 5492, 988.
\smallskip

\noindent Dye, S., Warren, S.J., Hambly, N.C., \et 2006,  MNRAS, 372, 1227.
\smallskip

\noindent Egan M P, Price S D, 1996 AJ 112, 2862
\smallskip

\noindent Fan, X., Strauss, M.A., Richards, G.T., \et 2001, AJ, 121, 31.
\smallskip

\noindent Foucaud, S., Almaini, O., Smail, I., Conselice, C.J., Lane, K.P., 
Edge, A.C., Simpson, C., Dunlop, J.S., McLure, R.J., Cirasuolo, M., Hirst, Hirst, P., 
Watson, M.G., Page, M.J., 2007, MNRAS, 376, L20.
\smallskip

\noindent Hambly, N. C., Hodgkin, S. T., Cossburn, M. R., Jameson, R. F., 1999, MNRAS, 303, 835.
\smallskip

\noindent Hambly, N. C., MacGillivray, H. T., Read, M. A., \et 2001 MNRAS 326, 1279.
\smallskip

\noindent Hambly, N.C., Read, M., Mann, R., Sutorius, E., Bond, I.,MacGillivray, H.,Williams, P.,Lawrence, A., 
2004a, ASP Conf.Proc. 314 137
\smallskip

\noindent Hambly, N.C., \et, 2004b, In: Optimizing Scientific Return for Astronomy
through Information Technologies, Proceedings of the SPIE, 5493, 423 (2004)
\smallskip

\noindent Hasegawa T., Oka T., Sato F., Tsuboi M., Yamazaki A. 1998, in `The central region of the Galaxy and galaxies', ed. Y.Sofue, Proc. IAU Symp. 184, p.171 (Dordrecht:Kluwer)
\smallskip

\noindent	Hawarden, T.G., Leggett, S. K., Letawsky, M.B., Ballantyne, D.R., Casali, M.M., 2001, 
MNRAS 325 563
\smallskip

\noindent Hewett, P.C., Warren, S.J., Leggett, S.K., Hodgkin, S.L., 2006 MNRAS, 367, 454.
\smallskip

\noindent Irwin M.J., 1985, MNRAS 214, 575.
\smallskip

\noindent Iovino, A., McCracken, H. J., Garilli, B.,  \et 2005, A\&A 442, 423
\smallskip

\noindent Jiang Z., Yao Y., Yang J., Ando M., Kato D., Kawai T., Kurita M., Nagata T.,
Nagayama T., Nakajima Y., Nagashima C., Sato S., Tamura M., Nakaya H., 
Sugitani K. 2002, AJ 577, 245
\smallskip

\noindent Knapp G.R.,Leggett S.K., Fan X. 2004, AJ 127 3553.
\smallskip

\noindent Kong, X., Daddi, E., Arimoto, N., \et 2006, ApJ 638 72.
\smallskip

\noindent Labb\'e I., \et, 2003, AJ, 125, 1107
\smallskip

\noindent Leggett, S. K., Geballe, T. R., Fan, X., \et 2000, ApJ, 536, L35.
\smallskip

\noindent Leggett, S.K.,  F. Allard, F., Burgasser, A.J., 2005, Proceedings of the 13th Cool Stars Workshop, ed. F.Favata, ESA Special Publications Series, ESA SP-560, p.143 (astroph/0409389).
\smallskip

\noindent Lodieu, N., Hambly, N.C., Jameson, R.F., Hodgkin, S.T., Carraro, G., 
Kendall, T.R., 2007, MNRAS, 374, 372.
\smallskip

\noindent Lopez-Corredoira, M., Cabrera-Lavers, A., Garz, F., Hammersley, P. L., 2002, A\&A, 394, 883
\smallskip

\noindent Mart\'{i}n, E.L., Magazz\`{u}, A., 2007, AN, in press
\smallskip

\noindent  McLure, R.J., Cirasuolo, M., Dunlop, J.S., Sekiguchi, K., Almaini, O., Foucaud, S., 
Simpson, C., Watson, M.G., Hirst, P., Page, M.J., Smail, I., 2006, MNRAS 372, 357.
\smallskip

\noindent Moraux, E., \et, 2007, AN, in press
\smallskip

\noindent  Nikolaev, S., Weinberg, M.D., Skrutskie, M.F., Cutri, R.M., Wheelock, S.L., Gizis, J.E., Howard, E.M.,	2000, AJ, 120, 3340.
\smallskip

\noindent Puget, P., Stadler, E., Doyon, R., \et 2004, SPIE, 5492, 978.
\smallskip

\noindent Robichon, N., Arenou, F., Turon, C., Mermilliod, J.C., 1999, 
A\&A, 345, 471
\smallskip

\noindent Sabbey, C.N., Coppi, P., Oemler, A., 1998, PASP, 110, 1067
\smallskip

\noindent Saracco, P., Giallongo, E., Cristiani, S., D'Odorico, S., Fontana, A.,
Iovino, A., Poli, F., Vanzella, E. 2001, A\&A 375, 1
\smallskip

\noindent Schneider D.~P., Gunn J.~E., Hoessel J.~G., 1983, ApJ, 264, 337
\smallskip

\noindent Simpson, S., Almaini, O., Cirasuolo, M., Dunlop, J., Foucaud, S., 
Hirst, P., Ivison, R., Page, M., Rawlings, S., Sekiguchi, K., Smail, I., 
Watson, M., 2006, MNRAS, 373, L21.
\smallskip

\noindent Skrutskie M.F., Cutri R.M., Stiening, R., \et 2006, AJ, 131, 1163.
\smallskip

\noindent Smith, G.P., Smail, I., Kneib, J.-P, \et, 2002, MNRAS, 330, 1.
\smallskip

\noindent Songaila., A., Cowie, L.L., Hu, E.M., Gardner, J.P., 1994 ApJS, 94, 461.
\smallskip

\noindent Ungerechts H, Thaddeus P, 1987 ApJS 63, 645
\smallskip

\noindent Venemans, B.P., McMahon, R.G., Warren, S.J., Gonzalez-Solares, E.A., 
Hewett, P.C., D. J. Mortlock, D.J., Dye, S., Sharp, R.G., 2007, MNRAS, 376, 76
\smallskip

\noindent Wang Q., Gotthelf E., \& Lang C. 2002, Nature, 415, 148
\smallskip

\noindent Warren, S.J., Hewett, P.C., Foltz, C.B, 2000, MNRAS, 312, 827
\smallskip

\noindent Warren, S.J., Lawrence, A., Almaini, O., \et, 2006, ESO Messenger, 126, 7
\smallskip

\noindent Warren, S.J., Hambly, N.C., Dye, S., \et 2007a, MNRAS, 375, 213.
\smallskip

\noindent Warren, S.J., Cross, N.J.G., Dye, S., \et 2007b, astro-ph/0703037.
\smallskip

\noindent Warren, S.J., Mortlock, D.J., Leggett, S.K., \et 2007c, MNRAS submitted.
\smallskip

\noindent York, D.G., Adelman, J., Anderson, J.E., \et 2000, AJ, 120, 1579
\smallskip

\noindent de Zeeuw, P.T., Hoogerwerf, R., de Bruijne, J.H.J., 1999, AJ, 117 354
\smallskip

\label{lastpage}

\end{document}